\documentclass[trackchanges, twocolumn]{aastex701}

\usepackage{xspace}

\newcommand{\ares}{SN\,Ares\xspace}
\newcommand{\cluster}{MACS J0308.9+2645}

\usepackage{amsmath}

\begin{document}

\title{SN Ares: A Strongly Lensed Type IIb Supernova at $\mathbf{z=1.276}$ with a Predicted 70-Year Time Delay}

\author[orcid=0000-0003-2037-4619]{Conor~Larison}
\email{conorjlarison@gmail.com}
\affiliation{Space Telescope Science Institute, 3700 San Martin Drive, Baltimore, MD 21218, USA}

\author[orcid=0000-0002-2361-7201]{Justin~D.~R.~Pierel}
\email{jpierel@stsci.edu}
\affiliation{Space Telescope Science Institute, 3700 San Martin Drive, Baltimore, MD 21218, USA}

\author[orcid=0000-0003-2445-3891]{Matthew Siebert}
\email{msiebert@stsci.edu}
\affiliation{Space Telescope Science Institute, 3700 San Martin Drive, Baltimore, MD 21218, USA}

\author[orcid=0009-0007-0522-7326]{Tom Resseguier}
\email{tresseg1@jhu.edu}
\affiliation{William H. Miller III Department of Physics \& Astronomy, Johns Hopkins University, 3400 N Charles St, Baltimore, MD 21218, USA}

\author[orcid=0000-0001-6278-032X]{Lukas J. Furtak}
\email{furtak@utexas.edu}
\affiliation{Cosmic Frontier Center, The University of Texas at Austin, Austin, TX 78712, USA}
\affiliation{Department of Astronomy, The University of Texas at Austin, Austin, TX 78712, USA}

\author[orcid=0000-0003-3484-399X]{Masamune Oguri}
\email{masamune.oguri@chiba-u.jp}
\affiliation{Center for Frontier Science, Chiba University, 1-33 Yayoi-cho, Inage-ku, Chiba 263-8522, Japan}
\affiliation{Department of Physics, Graduate School of Science, Chiba University, 1-33 Yayoi-Cho, Inage-Ku, Chiba 263-8522, Japan}

\author[orcid=0009-0007-2578-9238]{Gavin S. Farley}
\email{gavin.farley@mail.utoronto.ca}
\affiliation{David A. Dunlap Department of Astronomy and Astrophysics, University of Toronto, 50 St. George Street, Toronto, Ontario, M5S 3H4, Canada}


\author[orcid=0000-0003-3596-8794]{Hollis B. Akins}
\email{hollis.akins@gmail.com}
\altaffiliation{NASA Einstein Fellow}
\affiliation{Department of Astrophysical Sciences, Princeton University, 4 Ivy Lane, Princeton, NJ 08544, USA}

\author[orcid=0000-0003-3983-5438]{Yoshihisa Asada}
\email{yoshi.asada@utoronto.ca}
\affiliation{Dunlap Institute for Astronomy and Astrophysics, 50 St. George Street, Toronto, Ontario M5S 3H4, Canada}

\author[orcid=0000-0001-5063-8254]{Rachel Bezanson}
\email{rachel.bezanson@pitt.edu}
\affiliation{Department of Physics and Astronomy and PITT PACC, University of Pittsburgh, Pittsburgh, PA 15260, USA}

\author[orcid=0000-0001-5984-0395]{Maru\v{s}a Brada{\v c}}
\email{marusa.bradac@fmf.uni-lj.si}
\affiliation{University of Ljubljana, Faculty of Mathematics and Physics, Jadranska ulica 19, SI-1000 Ljubljana, Slovenia}
\affiliation{Department of Physics and Astronomy, University of California Davis, 1 Shields Avenue, Davis, CA 95616, USA}

\author[orcid=0000-0003-2680-005X]{Gabriel Brammer}
\email{gabriel.brammer@nbi.ku.dk}
\affiliation{Cosmic Dawn Center (DAWN), Denmark}
\affiliation{Niels Bohr Institute, University of Copenhagen, Jagtvej 128, DK-2200 Copenhagen N, Denmark}

\author[orcid=0000-0002-0302-2577]{John Chisholm}
\email{chisholm@austin.utexas.edu}
\affiliation{Department of Astronomy, The University of Texas at Austin, Austin, TX 78712, USA}

\author[orcid=0000-0001-7410-7669]{Dan Coe}
\email{dcoe@stsci.edu}
\affiliation{Space Telescope Science Institute (STScI), 3700 San Martin Drive, Baltimore, MD 21218, USA}
\affiliation{Center for Astrophysical Sciences, Department of Physics and Astronomy, The Johns Hopkins University, 3400 N Charles St. Baltimore, MD 21218, USA}
\affiliation{Association of Universities for Research in Astronomy (AURA), Inc.~for the European Space Agency (ESA)}

\author[orcid=0000-0001-8460-1564]{Pratika Dayal}
\email{pdayal@cita.utoronto.ca}
\affiliation{Canadian Institute for Theoretical Astrophysics, 60 St George St, University of Toronto, Toronto, ON M5S 3H8, Canada}
\affiliation{David A. Dunlap Department of Astronomy and Astrophysics, University of Toronto, 50 St. George Street, Toronto, Ontario, M5S 3H4, Canada}
\affiliation{Department of Physics, 60 St George St, University of Toronto, Toronto, ON M5S 3H8, Canada}

\author[orcid=0000-0001-7113-2738]{Henry C. Ferguson} \email{ferguson@stsci.edu}
\affiliation{Space Telescope Science Institute, 3700 San Martin Drive, Baltimore, MD 21218, USA}

\author[orcid=0000-0001-8519-1130]{Steven L. Finkelstein}
\email{stevenf@astro.as.utexas.edu}
\affiliation{Department of Astronomy, The University of Texas at Austin, Austin, TX 78712}
\affiliation{Cosmic Frontier Center, The University of Texas at Austin, Austin, TX 78712}

\author[orcid=0000-0001-7201-5066]{Seiji~Fujimoto}
\email{seiji.fujimoto@utoronto.ca}
\affiliation{David A. Dunlap Department of Astronomy and Astrophysics, University of Toronto, 50 St. George Street, Toronto, Ontario, M5S 3H4, Canada}
\affiliation{Dunlap Institute for Astronomy and Astrophysics, 50 St. George Street, Toronto, Ontario M5S 3H4, Canada}

\author[orcid=0009-0004-4332-9225]{Tomokazu Kiyota}
\email{tomokazu.kiyota@grad.nao.ac.jp}
\affiliation{Astronomical Science Program, Graduate Institute for Advanced Studies, SOKENDAI, 2-21-1 Osawa, Mitaka, Tokyo 181-8588, Japan}
\affiliation{National Astronomical Observatory of Japan, 2-21-1 Osawa, Mitaka, Tokyo, 181-8588, Japan}

\author[orcid=0000-0002-6610-2048]{Anton M. Koekemoer}
\email{koekemoer@stsci.edu}
\affiliation{Space Telescope Science Institute, 3700 San Martin Drive, Baltimore, MD 21218, USA}

\author[orcid=0000-0002-5588-9156]{Vasily Kokorev}
\email{vkokorev@utexas.edu}
\affiliation{Department of Astronomy, The University of Texas at Austin, Austin, TX 78712, USA}

\author{Minami Nakane}
\email{nakanem@icrr.u-tokyo.ac.jp}
\affiliation{Institute for Cosmic Ray Research, The University of Tokyo, 5-1-5 Kashiwanoha, Kashiwa, Chiba 277-8582, Japan}

\author[orcid=0000-0002-1049-6658]{Masami Ouchi}
\email{ouchims@icrr.u-tokyo.ac.jp}
\affiliation{National Astronomical Observatory of Japan, 2-21-1 Osawa, Mitaka, Tokyo 181-8588, Japan}
\affiliation{Institute for Cosmic Ray Research, The University of Tokyo, 5-1-5 Kashiwanoha, Kashiwa, Chiba 277-8582, Japan}
\affiliation{Department of Astronomical Science, SOKENDAI (The Graduate University for Advanced Studies), 2-21-1 Osawa, Mitaka, Tokyo, 181-8588, Japan}
\affiliation{Kavli Institute for the Physics and Mathematics of the Universe (WPI), The University of Tokyo, 5-1-5 Kashiwanoha, Kashiwa, Chiba 277-8583, Japan}

\author[orcid=0000-0002-9651-5716]{Richard Pan}
\email{richard.pan@tufts.edu}
\affiliation{Department of Physics \& Astronomy, Tufts University, Medford, MA 02155, USA}

\author[orcid=0000-0001-7503-8482]{Casey Papovich}
\email{papovich@tamu.edu}
\affiliation{Department of Physics and Astronomy, Texas A\&M University, 4242 TAMU, College Station, TX 77843-4242, USA}

\author[orcid=0000-0001-5492-1049]{Johan Richard}
\email{johan.richard@univ-lyon1.fr}
\affiliation{Universit\'e Claude Bernard Lyon 1, CRAL UMR5574, ENS de Lyon, CNRS, Villeurbanne, F-69622, France}

\author[orcid=0000-0002-6265-2675]{Luke Robbins}
\email{andrew.robbins@tufts.edu}
\affiliation{Department of Physics and Astronomy, Tufts University, 574 Boston Avenue, Suite 304, Medford, MA 02155, USA}

\author[orcid=0000-0002-4622-6617]{Fengwu Sun}
\email{fengwu.sun@cfa.harvard.edu}
\affiliation{Center for Astrophysics $|$ Harvard \& Smithsonian, 60 Garden St., Cambridge, MA 02138, USA}

\author{Kana Takechi}
\email{ktakechi@icrr.u-tokyo.ac.jp}
\affiliation{Institute for Cosmic Ray Research, The University of Tokyo, 5-1-5 Kashiwanoha, Kashiwa, Chiba 277-8582, Japan}

\author[orcid=0000-0002-8460-0390]{Tommaso Treu}
\email{tt@astro.ucla.edu}
\affiliation{Department of Physics \& Astronomy, University of California, Los Angeles, 430 Portola Plaza, Los Angeles, CA 90095, USA}

\author[orcid=0000-0002-4201-7367]{Chris Willott} \email{chris.willott@nrc.ca}
\affiliation{NRC Herzberg, 5071 West Saanich Rd, Victoria, BC V9E 2E7, Canada}

\author[orcid=0009-0006-6763-4245]{Hiroto Yanagisawa}
\email{yana@icrr.u-tokyo.ac.jp}
\affiliation{Institute for Cosmic Ray Research, The University of Tokyo, 5-1-5 Kashiwanoha, Kashiwa, Chiba 277-8582, Japan}
\affiliation{Department of Physics, Graduate School of Science, The University of Tokyo, 7-3-1 Hongo, Bunkyo, Tokyo 113-0033, Japan}


\author[orcid=0000-0002-5258-8761]{Abdurro'uf}
\email{abdurroufastro@gmail.com}
\affiliation{Department of Astronomy, Indiana University, 727 East Third Street, Bloomington, IN 47405, USA}

\author[orcid=0000-0003-2718-8640]{Joseph~F.~V.~Allingham}
\email{allingha@post.bgu.ac.il}
\affiliation{Department of Physics, Ben-Gurion University of the Negev, P.O. Box 653, Be'er-Sheva 84105, Israel}

\author[orcid=0000-0001-5758-1000]{Ricardo O. Amor\'{i}n} \email{amorin@iaa.es}
\affiliation{Instituto de Astrof\'{i}sica de Andaluc\'{i}a (CSIC), Apartado 3004, 18080 Granada, Spain}

\author[orcid=0000-0002-8686-8737]{Franz E. Bauer}
\email{franz.e.bauer@gmail.com}
\affiliation{Instituto de Alta Investigaci{\'{o}}n, Universidad de Tarapac{\'{a}}, Casilla 7D, Arica, 1010000, Chile}

\author[orcid=0000-0002-7908-9284]{Larry Bradley}
\email{lbradley@stsci.edu}
\affiliation{Space Telescope Science Institute, 3700 San Martin Drive, Baltimore, MD 21218, USA}

\author[orcid=0000-0002-1537-6911]{Mateusz Bronikowski}
\email{bronikowski.mateusz@gmail.com}
\affiliation{Astronomical Observatory, University of Warsaw, Al. Ujazdowskie 4, 00-478 Warszawa, Poland}

\author[orcid=0000-0003-1949-7638]{Christopher J. Conselice}
\email{conselice@manchester.ac.uk}
\affiliation{Jodrell Bank Centre for Astrophysics, University of Manchester, Oxford Road, Manchester}

\author[orcid=0000-0003-4263-2228]{David~A.~Coulter}
\email{dcoulter@stsci.edu}
\affiliation{Physics and Astronomy Department, Johns Hopkins University, Baltimore, MD 21218, USA}
\affiliation{Space Telescope Science Institute, Baltimore, MD 21218, USA}

\author[orcid=0000-0002-7566-6080]{James~M.~DerKacy}
\email{jderkacy@stsci.edu}
\affiliation{Space Telescope Science Institute, 3700 San Martin Drive, Baltimore, MD 21218, USA}

\author[orcid=0000-0001-9065-3926]{Jos\'{e}~M.~Diego}
\email{jdiego@ifca.es}
\affiliation{Instituto de F\'isica de Cantabria (CSIC-UC), Avda. Los Castros s/n., 39005 Santander, Spain}

\author[orcid=0000-0003-1344-9475]{Eiichi Egami}
\email{egami@arizona.edu}
\affiliation{Steward Observatory, University of Arizona, 933 N Cherry Ave, Tucson, AZ 85721, USA}

\author[orcid=0000-0002-9382-9832]{Andreas~L.~Faisst}
\email{afaisst@caltech.edu}
\affiliation{IPAC, California Institute of Technology, 1200 E. California Blvd. Pasadena, CA 91125, USA}

\author[orcid=0000-0003-2238-1572]{Ori D. Fox}
\email{ofox@stsci.edu}
\affiliation{Space Telescope Science Institute, 3700 San Martin Drive, Baltimore, MD 21218, USA}

\author[orcid=0000-0003-1625-8009]{Brenda~L.~Frye}
\email{bfrye@arizona.edu}
\affiliation{Department of Astronomy/Steward Observatory, 933 N. Cherry Ave., Tucson, AZ 85721}

\author[orcid=0000-0001-7440-8832]{Yoshinobu Fudamoto}
\email{yoshinobu.fudamoto@gmail.com}
\affiliation{Center for Frontier Science, Chiba University, 1-33 Yayoi-cho, Inage-ku, Chiba 263-8522, Japan}

\author[orcid=0000-0002-4837-1615]{Mauro Gonz\'{a}lez-Otero}
\email{mauromarago@gmail.com}
\affiliation{Instituto de Astrof\'isica de Andaluc\'ia--CSIC, Glorieta de la Astronom\'ia s/n, E--18008 Granada, Spain}

\author[orcid=0000-0002-5060-1379]{Massimo~Griggio}
\email{mgriggio@stsci.edu}
\affiliation{Space Telescope Science Institute, 3700 San Martin Drive, Baltimore, MD 21218, USA}

\author[orcid=0000-0001-6145-5090]{Nimish P. Hathi}
\email{nphathi@gmail.com}
\affiliation{U.S. National Science Foundation, Alexandria, VA, USA}

\author[orcid=0000-0003-4512-8705]{Tiger Yu-Yang Hsiao}  
\email{tiger.hsiao@utexas.edu}
\affiliation{Department of Astronomy, The University of Texas at Austin, 2515 Speedway, Stop C1400, Austin, TX 78712, USA} 
\affiliation{Cosmic Frontier Center, The University of Texas at Austin, Austin, TX 78712, USA}

\author[orcid=0000-0001-8738-6011]{Saurabh~W.~Jha}
\email{saurabh@physics.rutgers.edu}
\affiliation{Department of Physics and Astronomy, Rutgers University, Piscataway, NJ 08854, USA}

\author[orcid=0000-0002-6090-2853]{Y. Jim\'enez-Teja}
\email{yojite@iaa.es}
\affiliation{Instituto de Astrof\'isica de Andaluc\'ia--CSIC, Glorieta de la Astronom\'ia s/n, E--18008 Granada, Spain}
\affiliation{Observat\'orio Nacional, Rua General Jos\'e Cristino, 77 - Bairro Imperial de S\~ao Crist\'ov\~ao, Rio de Janeiro, 20921-400, Brazil}

\author[orcid=0000-0001-9187-3605]{Jeyhan S. Kartaltepe}
\email{jeyhan@astro.rit.edu}
\affiliation{Laboratory for Multiwavelength Astrophysics, School of Physics and Astronomy, Rochester Institute of Technology, 84 Lomb Memorial Drive, Rochester, NY 14623, USA}

\author[orcid=0000-0002-3475-7648]{Gourav Khullar}
\email{gkhullar@uw.edu}
\affiliation{Department of Astronomy \& the DiRAC Institute, University of Washington, Physics-Astronomy Building, Box 351580, Seattle, WA 98195-1700, USA}

\author[0009-0003-8380-4003]{Zachary~G.~Lane}
\email{zalane@stsci.edu}
\affiliation{Space Telescope Science Institute, 3700 San Martin Drive, Baltimore, MD 21218, USA}

\author[orcid=0000-0003-2540-7424]{Paulo A. A. Lopes}
\email{plopes@ov.ufrj.br}
\affiliation{Observat\'{o}rio do Valongo, Universidade Federal do Rio de Janeiro, Ladeira do Pedro Ant\^{o}nio 43, Rio de Janeiro RJ 20080-090, Brazil}

\author[orcid=0000-0003-1581-7825]{Ray A. Lucas}
\email{lucas@stsci.edu}
\affiliation{Space Telescope Science Institute, 3700 San Martin Drive, Baltimore, MD 21218, USA}

\author[orcid=0000-0002-4872-2294]{Georgios~E.~Magdis}
\email{geoma@dtu.dk}
\affiliation{Cosmic Dawn Center (DAWN), Denmark}
\affiliation{DTU-Space, Elektrovej, Building 328, 2800, Kgs. Lyngby, Denmark}

\author[orcid=0000-0003-1427-2456]{Matteo Messa}
\email{matteo.messa@inaf.it}
\affiliation{INAF -- OAS, Osservatorio di Astrofisica e Scienza dello Spazio di Bologna, via Gobetti 93/3, I-40129 Bologna, Italy}

\author[orcid=0000-0003-3729-1684]{Rohan P. Naidu}
\email{rnaidu@hawaii.edu}
\affiliation{Institute for Astronomy, University of Hawai`i at M\={a}noa, Honolulu, HI, USA}

\author[orcid=0000-0002-6150-833X]{Rafael Ortiz~III}
\email{rortizii@asu.edu}
\affiliation{School of Earth and Space Exploration, Arizona State University, Tempe, AZ 85287-1404, USA}

\author[orcid=0000-0002-2282-8795]{Massimo~Pascale}
\email{mpascale@astro.ucla.edu}
\affiliation{Department of Physics \& Astronomy, University of California, Los Angeles, 430 Portola Plaza, Los Angeles, CA 90095, USA}

\author[orcid=0000-0002-4410-5387]{Armin~Rest}
\email{arest@stsci.edu}
\affiliation{Space Telescope Science Institute, 3700 San Martin Drive, Baltimore, MD 21218, USA}
\affiliation{William H. Miller III Department of Physics \& Astronomy, Johns Hopkins University, 3400 N Charles St, Baltimore, MD 21218, USA}

\author[orcid=0000-0003-4223-7324]{Massimo Ricotti}
\email{ricotti@umd.edu}
\affiliation{Department of Astronomy, University of Maryland, College Park, 20742, USA}

\author{Melissa~Shahbandeh}
\email{mshahbandeh@stsci.edu}
\affiliation{Space Telescope Science Institute, 3700 San Martin Drive, Baltimore, MD 21218, USA}

\author[orcid=0000-0003-1889-0227]{William Sheu}
\email{wsheu@astro.ucla.edu}
\affiliation{Department of Physics \& Astronomy, University of California, Los Angeles, Los Angeles, CA 90095, USA}

\author[orcid=0000-0002-2798-2943]{Koji~Shukawa}
\email{kshukawa@stsci.edu}
\affiliation{William H. Miller III Department of Physics \& Astronomy, Johns Hopkins University, 3400 N Charles St, Baltimore, MD 21218, USA}
\affiliation{Space Telescope Science Institute, 3700 San Martin Drive, Baltimore, MD 21218, USA}

\author[orcid=0000-0001-8156-6281]{Rogier~A.~Windhorst}
\email{Rogier.Windhorst@asu.edu}
\affiliation{School of Earth \& Space Exploration, Arizona State University, Tempe, AZ 85287-6004, USA}

\author[orcid=0000-0001-8600-7008]{Anita~Zanella}
\email{anita.zanella@inaf.it}
\affiliation{Istituto Nazionale di Astrofisica (INAF), Via Gobetti 93/3, 40129, Bologna, Italy}

\author[orcid=0000-0002-0632-8897]{Yossef Zenati}
\email{yzenati1@jhu.edu}
\affiliation{Astrophysics Research Center of the Open University (ARCO), Department of Natural Sciences, Ra'anana 4353701, Israel}
\affiliation{William H. Miller III Department of Physics \& Astronomy, Johns Hopkins University, 3400 N Charles St, Baltimore, MD 21218, USA}

\author[orcid=0000-0002-0350-4488]{Adi Zitrin}
\email{adizitrin@gmail.com}
\affiliation{Department of Physics, Ben-Gurion University of the Negev, P.O. Box 653, Be'er-Sheva 84105, Israel}

\begin{abstract}

We present the discovery of supernova (SN)~Ares, a strongly lensed supernova at $z_{\rm spec}=1.276\pm0.002$ in a massive, triply imaged host galaxy behind the \cluster\ galaxy cluster. \textit{JWST}/NIRSpec spectroscopy spanning the rest-frame optical to NIR of Ares reveals helium features at 5876 \AA, 7065 \AA, and 10830 \AA\, as well as H$\alpha$ and Pa$\beta$. A comparison to local core-collapse SN spectra and light curves leads to a classification of \ares\ as a Type IIb supernova, the highest-redshift member of its class spectroscopically identified. A comparison to local SN IIb light curves provides a phase estimate for the spectrum of $34_{-9}^{+10}$ rest-frame days post-radioactive peak, providing a unique view into a high-redshift stripped-envelope supernova in its late-photospheric evolution. A lens model of the cluster constructed using recent \textit{JWST} data indicates that \ares\ is the first image to arrive and will reappear in two future images near the center of the cluster plane in $\gtrsim 70$ years, with predicted magnifications of $\sim5$x and $\sim1.5$x. Planned \textit{JWST}/NIRSpec Multi-Object Spectroscopy of the cluster will allow for more detailed modeling of the lensed system for tighter constraints on the time delay and magnifications, as well as a template spectrum at the location of \ares.

\end{abstract}




\section{Introduction} \label{sec:intro} 

The phenomenon of strong gravitational lensing occurs when a massive foreground object, such as a galaxy or galaxy cluster, distorts the light from a background source. This results in multiple magnified images appearing at different times (for reviews on galaxy- and cluster-scale strong lensing, see \citealt{Shajib_review_2024} and \citealt{Natarajan_review_2024}). The offset in arrival time between images can be anywhere from hours to decades depending on the mass of the foreground lens and the geometry of the system, with cluster-scale lenses typically producing longer time delays \citep{Oguri_review_2019}. The absolute and relative size of these delays is related to the physical scale of the Universe through the geometric configuration of the lens and background source \citep{linder_lensing_2011}.

Due to the direct relationship between these delays and geometric distance, a measurement of the offset in image arrival times leads to constraints on cosmological parameters such as the local expansion rate of the Universe, $H_0$\citep{refsdal_possibility_1964}. The use of these delays for cosmological measurements is known as ``time-delay cosmography'' \citep{refsdal_possibility_1964,treu_time_2016,treu2026} and has found great success with galaxy-scale lensed quasar systems \citep{wong_h0licow_2020,tdcosmo_collaboration_tdcosmo_2025,TDCOSMO26,TDCOSMO26L}. Recently, strongly lensed supernovae (SNe) have emerged as incredible tools for time-delay cosmography as well. So far, four measurements of $H_0$ have been made with strongly lensed SNe (Refsdal, H0pe, Encore, and 2025wny), finding $H_0$ values that span the current cosmological tension, from $65 \,\rm{kms^{-1}Mpc^{-1}}$ to $76 \,\rm{kms^{-1}Mpc^{-1}}$ \citep{kelly_constraints_2023,grillo_cosmography_2024,pascale_sn_2025,pierel_cosmology_2025,SuyuEncore2026,Bazzanini:2026,Johansson_wny_h0_2026,Mortsell_wny_2026,Schuldt_encore_2026}. The advantages of lensed SN systems in particular are clear: the lensed SNe have predictable light curves with distinct features (such as a radioactive peak) that simplify time-delay measurements; they require short observing campaigns of weeks to months; they are resistant to the effects of chromatic microlensing \citep{goldstein_precise_2018,pierel_turning_2019,huber_holismokes_2021}, and they fade on timescales which allow for template observations after the SN has faded \citep{Ding:2021}. Regardless of whether the source is a lensed SN or quasar, cluster-scale systems possess an advantage over galaxy-scale systems in that they include multiple background sources across redshifts, which helps alleviate the mass sheet degeneracy uncertainty at the price of a more complex model \citep{BirrerMassSheet2024}. Conversely, clusters are not dynamically relaxed systems. As such, traditional dynamical methods cannot be used to help constrain the multi-peaked mass model and reduce mass-sheet degeneracy. Furthermore, cluster environments and line of sight objects which contribute to the lensing potential are likely to be complex, requiring a dedicated analysis. Finally, for simply-parametrized models, residual errors in predicting image positions are of order tenths of arcseconds \citep{Bazzanini:2026}, indicating that structures with Einstein radii smaller than this are not fully accounted for. In order to better understand their relative strengths and weaknesses, it is crucial to continue building samples of strongly lensed SNe at both the galaxy and cluster scales.

\begin{figure*}
    \centering
    \includegraphics[width=\linewidth]{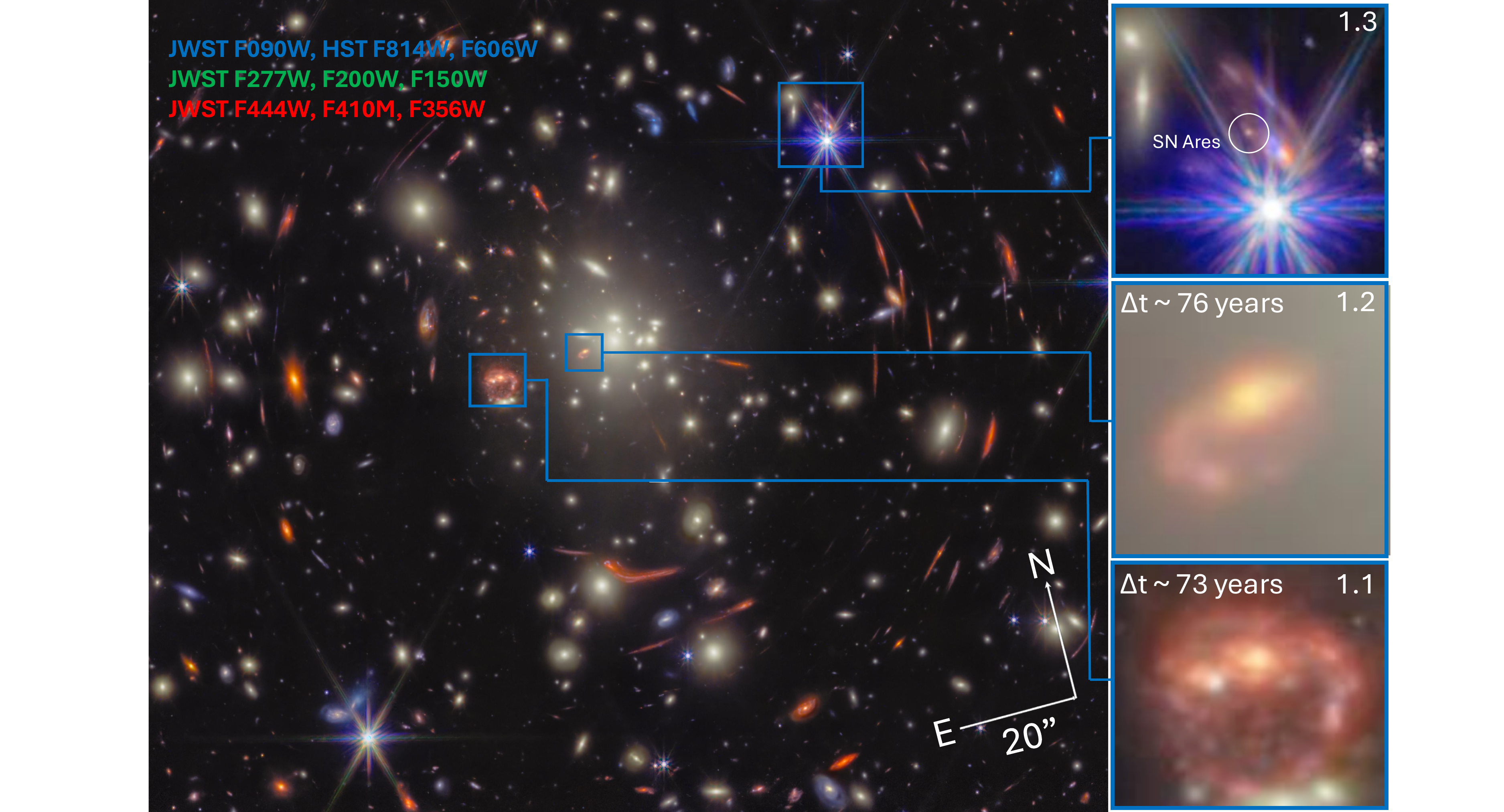}
    \caption{Color image of \cluster\ showing the detection of SN Ares in its multiply imaged host galaxy. A label describing the filters that contribute to each color channel are shown on the upper left, while orientation and scale of the image is shown on the bottom right. Cutouts are provided for the three images of the SN Ares host galaxy, with the predicted time delays to images 1.1 and 1.2 labeled. Image 1.3 contains the discovered image of \ares, highlighting its large star-forming host and the bright, nearby foreground star.
    }
    \label{fig:panchromatic}
\end{figure*}

Apart from their usefulness for cosmology, strongly lensed SNe also allow for unprecedented studies of high-redshift SN physics. Such studies elucidate the evolution in SN and SN progenitor properties, as well as inform us on how chemical enrichment has occurred over cosmic time. Without the magnification afforded by lensing, however, it would be impossible to detect (much less comprehensively analyze) certain classes of SNe out to high redshifts. For example, SN Refsdal (a 1987A-like core-collapse SN at $z=1.49$), would have had a peak magnitude $m_V > 27$ mag and would have been undetectable by \textit{HST} without its $\mu=15$ magnification \citep{kelly_sn_2016}. Furthermore, it would have been impossible for \textit{HST} to have obtained a spectrum that could even classify this event at all. Similarly, the recently discovered Type II SN Eos at $z=5.13$, would have been completely undetectable without the $\sim30$x boost ($\sim 4$ magnitudes) in brightness afforded by lensing \citep{Allingham_mj1931_2026,Coulter_eos_2026,Asada2026}. Not only did lensing allow for its discovery, a remarkably high signal-to-noise spectrum obtained with \textit{JWST}/NIRSpec allowed for a direct measurement of the Fe II 5018 \AA\ absorption component, directly confirming the expectation based on its host galaxy that it was formed in a low-metallicity environment. Another recent strongly lensed SN at $z\approx2$, 2025wny, a member of an intrinsically bright sub-class known as superluminous SNe \citep{Gal-Yam_slsne_2019}, was also magnified by $\mu\gtrsim20$, allowing for high signal-to-noise spectroscopy from multiple ground based observatories in addition to space-based resources \citep{johansson_discovery_2025,Goobar2026,LiWny2026}. Although this SN is intrinsically bright ($M_{\rm UV} \lesssim -21$ mag), the massive boost by lensing allows for a much more detailed study than would be possible otherwise. These recent discoveries show how strongly lensed SNe can be simultaneously valuable for precision time-delay cosmography as well as studies of high-redshift SN physics, greatly extending the parameter space in terms of SN type and brightness for which we can make spectroscopic observations. Although we can expect many more spectroscopically confirmed SNe in the field with \textit{Roman}, studies such as these that probe fundamental spectroscopic properties at high redshifts will only remain possible with \textit{JWST} observations of lensed SNe.

Here, we present the discovery and \textit{JWST} + \textit{HST} follow-up of SN Ares, a strongly lensed, core-collapse SN at $z=1.276$. In Section~\ref{sec:obs}, we present the spectrophotometric dataset of SN Ares. The lens model analysis of \cluster\ and the resulting $\sim70$-year reapperances of \ares\ is discussed in Section~\ref{sec:lens_model}. In Section~\ref{sec:spec_class}, we show the \textit{JWST}/NIRSpec FS observation of SN Ares and the resulting classification as an SN IIb. We combine this classification with our photometric dataset to form our best estimate for the phase of the SN in Section~\ref{sec:phase}. Finally, in Section~\ref{sec:discussion} we conclude about the future possible use of SN~Ares as an incredibly precise long-baseline time delay experiment, as well as the takeaways for high-redshift SN physics.

\begin{figure*}
    \centering
    \includegraphics[width=\linewidth]{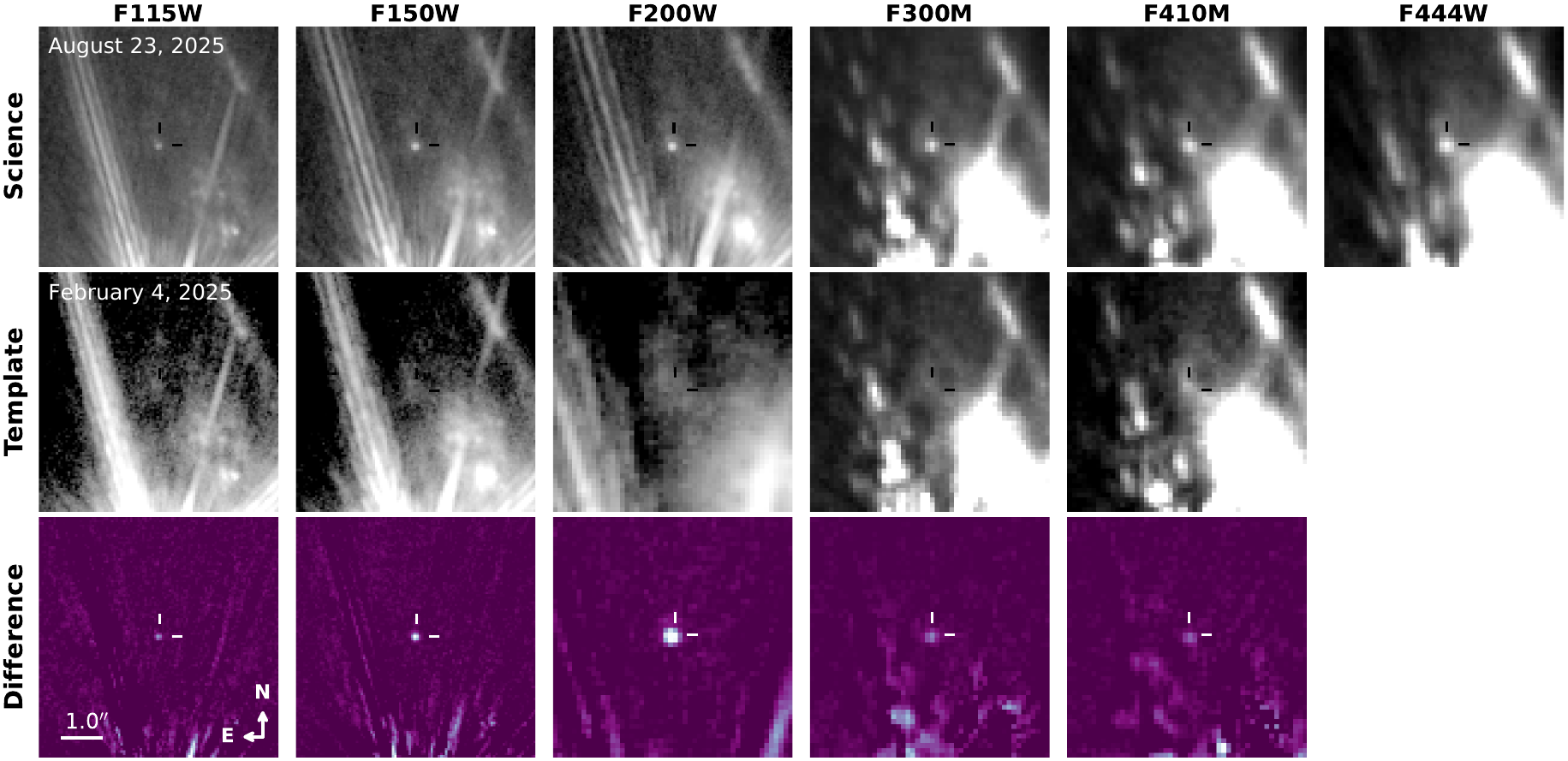}
    \caption{Discovery-epoch \textit{JWST} imaging of SN Ares, with the science images shown at the top, template images in the middle, and differences on the bottom. A compass and angular scale marker are shown on the bottom left panel. Due to the high-resolution imaging, contamination by the bright foreground star is greatly mitigated. Template images were unavailable for the F444W image, so only the science image panel is shown.}
    \label{fig:diffs}
\end{figure*}

\section{Observations \label{sec:obs}}

\ares\ ($\alpha = 03^{\mathrm{h}} 08^{\mathrm{m}} 53^{\mathrm{s}}_{^{\centerdot}}418$, $\delta = +26^\circ45' 58\farcs63$, J2000) was discovered in {\it JWST}/NIRCam imaging acquired on 2025 August 23 (UTC dates are used throughout, MJD times of observations found in Table~\ref{tab:ares_photometry}) by the Vast Exploration for Nascent, Unexplored Sources \citep[VENUS, Cycle 4 GO: PID 6882;][]{Fujimoto_venus_2025} collaboration in the \cluster\ cluster field behind a bright foreground star in a large star-forming and multiply imaged host. A color image of this field constructed with \textit{HST} and \textit{JWST} data is shown in Figure~\ref{fig:panchromatic}.

\subsection{JWST Data}

The \textit{JWST}/NIRCam discovery data of \ares\ taken by VENUS consists of ten filters: (F090W, F115W, F150W, F200W, F210M, F277W, F300M, F356W, F410M, amd F444W). NIRCam images in the F115W, F150W, F200W, F250M, F300M, and F410M filters were taken prior to the explosion of \ares\ on 2025 February 4 \citep[Cycle 3 GO: PID 5293;][]{xu_jwst_24,abdurrouf_mj0308_2025}, which are used as template images for our analysis. The imaging is reduced with the standard VENUS \texttt{grizli} infrastructure \citep{brammer2023}, used by all VENUS studies \citep[e.g.,][]{nakane2025, golubchik2026, coulter2026eos, yanagisawa2026, allingham26, asada2026eos}, which applies the level-2 detector calibrations with corrections for snowball artifacts, wisp features, and $1/f$ noise. This pipeline also aligns the images to the Gaia~DR3 frame and produces level-3 drizzled images (whereby all dithered exposures are combined into a single mosaic) on a common reference frame, thus enabling robust difference imaging. Cutouts of the science, template, and difference images from this pipeline procedure are presented in Figure~\ref{fig:diffs}. We performed aperture photometry with local background subtraction and aperture corrections using the \texttt{space\_phot} package \citep{pierel_space-phot_2024}, the results of which are shown in Table~\ref{tab:ares_photometry}.


Following the discovery of \ares, a \textit{JWST} Director's Discretionary Time program was approved to obtain an epoch of F115W/F444W NIRCam imaging and a NIRSpec spectrum (\textit{JWST} DDT PID: 9478). These data were obtained on 2025 September 23, an observer-frame month from the discovery epoch. NIRSpec fixed-slit PRISM/CLEAR spectroscopy of SN~Ares was obtained through the \texttt{S200A1} aperture in a five-point nod pattern, for a total exposure time of 4012~s. We retrieved the stage-1 products from MAST, where they were processed with the \textit{JWST}\footnote{\url{https://github.com/spacetelescope/jwst}} pipeline \citep[version 1.20.2;][]{Bushouse_JWST_Calibration_Pipeline_2024} routines for saturation flagging, bias and dark subtraction, jump detection, and ramp fitting.

We reduced these products with \texttt{msaexp}\footnote{\url{https://github.com/gbrammer/msaexp}} \citep[version 0.6.17;][]{Brammer23}, which calls the \textit{JWST} pipeline routines for wavelength calibration, slit rectification, flat-field correction, and flux calibration. We constructed a sky background model by fitting a cubic spline in wavelength to slit pixels unaffected by SN and host galaxy light. The five exposures were then drizzle-combined onto a common rectified grid, and we performed an optimal point-source extraction. Finally, we applied an additional local background subtraction to remove residual background emission from the host galaxy. The reduced spectrum is shown in Figure \ref{fig:jwst_spec}.

\subsection{HST Data}

Further \textit{HST}/WFC3 UVIS/IR observations in the F814W, F110W, and F160W filters of the \ares\ field were obtained following the \textit{JWST} discovery \citep[\textit{HST} GO PID: 17474,][]{Pierel_prop_2023}. Due to a combination of the lower resolution of \textit{HST} and contamination from the nearby foreground star, these data produced only upper limits on the SN Ares light curve evolution. We present a discussion of these data and present upper limits for F160W in Appendix~\ref{app:hst_upper}. Future template observations matching more closely to the position angle of our observations may allow for a detection, however, such observations are currently not planned.

\begin{deluxetable}{lcc}
\tablecaption{\textit{JWST}/NIRCam photometry of SN~Ares.
\label{tab:ares_photometry}}
\tablewidth{\columnwidth}
\tabletypesize{\small}
\tablehead{
\colhead{MJD} &
\colhead{Filter} &
\colhead{$m_{\rm AB}$ (mag)}
}
\startdata
60910.6806 & F115W & $26.793 \pm 0.097$ \\[3pt]
60910.7067 & F150W & $25.641 \pm 0.032$ \\[3pt]
60910.7286 & F200W & $25.453 \pm 0.021$ \\[3pt]
60910.7514 & F300M & $25.600 \pm 0.035$ \\[6pt]
60910.6503 & F410M & $25.872 \pm 0.058$ \\[3pt]
60910.6806 & F444W & $25.639 \pm 0.024$ \\[3pt]
60941.2733 & F115W & $27.297 \pm 0.154$ \\[3pt]
60941.2733 & F444W & $26.007 \pm 0.030$
\enddata
\end{deluxetable}

\section{Lens model and predictions} \label{sec:lens_model}

In order to estimate the uncertainty in future arrival times of the SN Ares images, we rely on three independent lens models of \cluster. We note that these lens model predictions will be greatly improved when we take scheduled \textit{JWST}/NIRSpec Multi-Object Spectroscopcy (MOS) observations of the field in early 2027.

The first model uses a collection of dual Pseudo Isothermal Elliptical mass distributions (dPIE) using \texttt{Lenstool} \citep{lenstool_2011}. Cluster members were identified using a double red sequence method and the lens model is constrained by 48 multiple images belonging to 15 independent families. Two of these families have spectroscopic redshifts while the rest rely on tight photometric redshifts from the 17 combined \textit{HST} and \textit{JWST} filters. The lens model has an RMS of $0.''66$ and was used to estimate $\mu=5.6 \pm 0.5$ for SN Ares. According to the lens model, we are observing the first of three arrivals of SN Ares. We note that the arrival order is defined by the geometry of the lensing system and is not sensitive to a particular model. With $H_0=70\,\rm{km\,s^{-1}Mpc^{-1}}$, the second arrival is predicted in $70\pm1$ years with $\mu=4.9\pm0.1$, while the third and last arrival is predicted in $73\pm1$ years with $\mu=1.5\pm0.1$. A difference of 3 $\rm{kms^{-1}Mpc^{-1}}$ in the value of $H_0$ leads roughly to a difference of 5 years in the time delays (with a lower value of $H_0$ leading to a higher time delay). A final monitoring campaign must cover both uncertainty from the lens models and uncertainty in the cosmological model. Over time, as both lens model methods and constraints on \cluster\ improve, both sources of uncertainty will certainly be better constrained, thus tightening the window over which the cluster must be monitored. A detailed description of the method used to construct this lens model will be presented in Resseguier et al. (in prep.).

Another strong lens mass model constructed with the \textsc{glafic} software \citep{oguri2010,oguri2021b} predicts similar magnifications and time delays, including a magnification of $\mu=4.7\pm0.3$ for SN Ares, a second arrival in $72 \pm 2 $ years, and a third arrival in $77 \pm 4$ years, again assuming $H_0=70\,\rm{km\,s^{-1}Mpc^{-1}}$. The lens model will be presented in M. Oguri et al. (in prep.).

A third parametric, \textit{JWST}-based SL mass model of \cluster\ was derived with the code \texttt{AstroLensPy} \citep[\texttt{ALP};][]{allingham26}, which is a \texttt{python} implementation of the \citet{zitrin2015} analytic method. It is constrained with 45 multiple images of 21 sources, two of which have spectroscopic redshifts, and achieves a lens-plane average image reproduction error $\Delta_{\mathrm{RMS}}=0.37''$. According to this model, SN Ares is magnified by a factor $\mu=4.6\pm0.4$, in agreement with both the \texttt{lenstool} and \texttt{glafic} models at $1\sigma$, and will re-appear in $76.4\pm0.3$\,yr, and again in $79.1\pm0.4$\,yr. This model will be published in Allingham et al.\ (in prep.).

\begin{figure*}
    \centering
    \includegraphics[width=\linewidth]{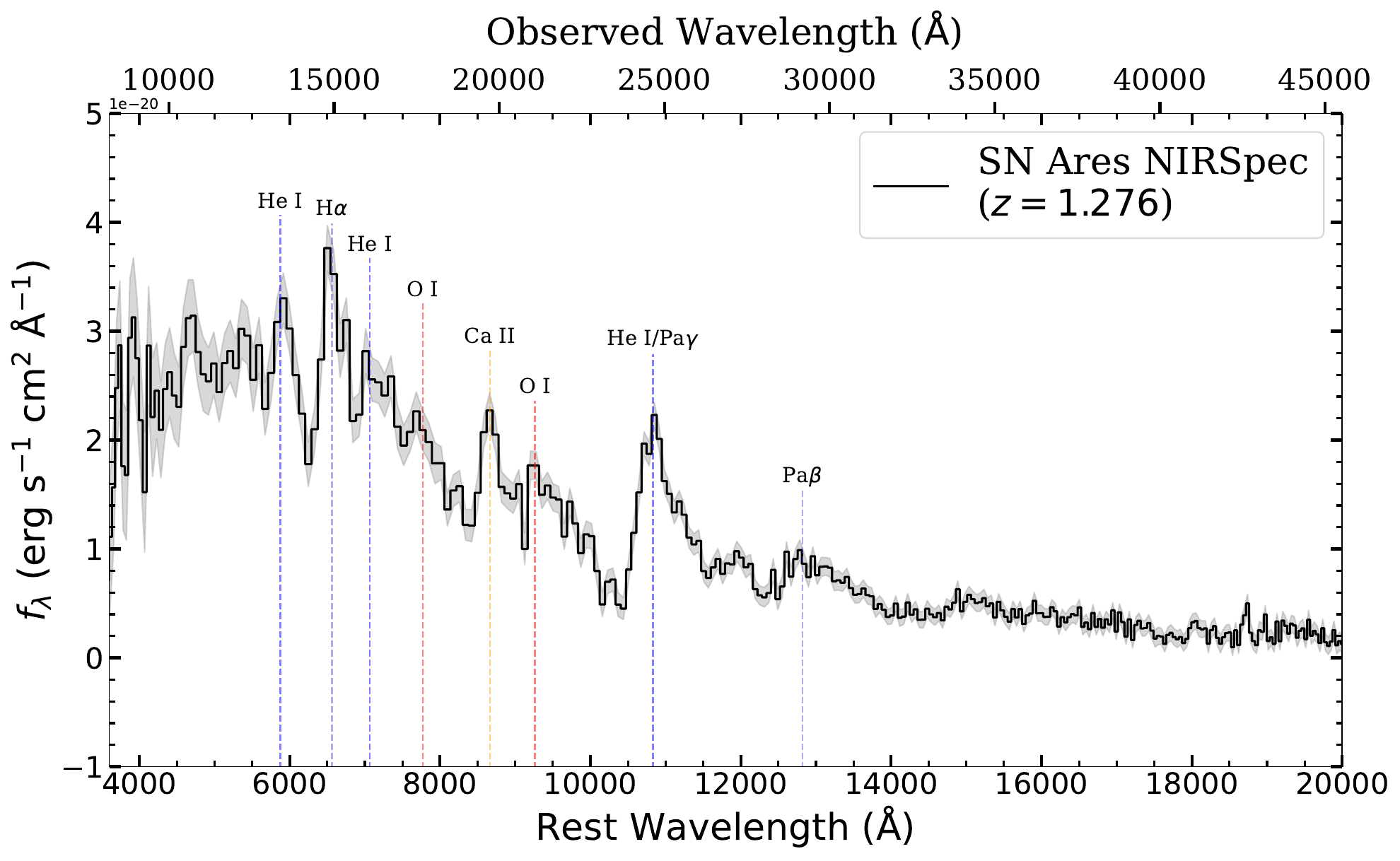}
    \caption{\textit{JWST}/NIRSpec spectrum of \ares\ with $1\sigma$ uncertainties overplotted. We show line identifications for the diagnostic features, including the He I $5876$, $7065$, and $10830$~\AA\ lines, H$\alpha$, and Pa$\beta$. The $5876$ and $10830$~\AA\ helium lines are detected at high significance, while we find a $2.6 - 3.2\sigma$ detection of the $7065$~\AA\ He line depending on the definition of the continuum region. We note that there remains substantial host contribution to the blue continuum and likely to the H$\alpha$ feature; however, a weaker hydrogen feature would even more strongly favor an SN~IIb classification.}
    \label{fig:jwst_spec}
\end{figure*}

\section{Spectroscopic classification \label{sec:spec_class}}

In Figure~\ref{fig:jwst_spec} we show the NIRSpec spectrum of \ares\ from rest-frame $4000$ -- $20000$ \AA, where there is sufficient signal-to-noise for spectroscopic comparison. We measured the redshift for
SN Ares from a 1D spectrum extracted with a boxcar aperture centered on the host galaxy. We modeled its continuum and spectral lines with additional routines in \texttt{msaexp} \citep{Brammer23}. Several strong emission lines fix the redshift, including [O II] $\lambda3727$, H$\beta$ + [O III] $\lambda5007$, H$\alpha$ +[N II] $\lambda \lambda$6548,6583, [S II] $\lambda \lambda$6717,6731, He I $\lambda$10830, Pa$\beta$, and Pa$\alpha$. The best fit is z = $1.276 \pm 0.002$. We show the spectrum used for this fit in Appendix~\ref{app:redshift}. For classification, we favor the Next Generation SuperFit (NGSF) code, as it fits a host continuum contribution to the spectrum jointly with the SN templates \citep{Howell_superfit_2005,Goldwasser:2022}. This is useful in the case of SN Ares, which has host contribution in its spectrum even after our best local host subtraction. From this, we found that the best classification is that of a Type IIb SN (SN IIb), a partially stripped massive star explosion with a far smaller hydrogen envelope than nonstripped core-collapse SN progenitors. This smaller envelope results in an early spectroscopic evolution similar to SNe IIP/L which then transitions to a spectroscopic series more similar to SNe Ib/c \citep{Branch_wheeler_2017}. Since H$\alpha$ is detected, it is immediately apparent that \ares\ is not an SN Ib/c, which lack signatures of hydrogen entirely.

The main diagnostic that favors \ares\ as an SN IIb rather than a SN IIP/L, is the existence and strength of the helium features, including the He I $5876$~\AA\ and $7065$~\AA\ features that lie to either side of the H$\alpha$ profile. The existence of lines at these wavelengths is not necessarily an immediate indication of an SN IIb classification, as both may appear at later times for SNe~IIP/L, with a Na I doublet also contributing to the blue emission feature. However, the large ratio of the $5876$~\AA\ feature to H$\alpha$ strength strongly favors an SN IIb classification, rather than that of an SN IIP/L. Even in late times, the He I (mixed with Na I) $5876$~\AA\ and He I $7065$~\AA\ features are much weaker relative to H$\alpha$ than we observe for \ares. The same argument is true for the size of the He I $10830$ \AA\ to Pa$\beta$ line ratio, which is much higher for \ares\ than would be expected of a typical SN II, except perhaps at nebular times. And if \ares\ were a late-time SN~IIP/L, there would be other emergent forbidden lines ([O I], [Ca II], [Fe II]) that are absent from the spectrum.

\begin{table}[t]
\centering
\caption{Type IIb spectra from the OSC used in the NGSF analysis.}
\label{tab:osc_iib_spectra}
\small
\setlength{\tabcolsep}{4pt}
\begin{tabular}{lrl}
\hline
Supernova & $N_{\rm spec}$ & References \\
\hline
SN 1993J  & 99 & [1, 2, 3, 4, 5] \\
SN 1996cb & 29 & [4, 5, 6] \\
SN 2001ig & 15 & [7] \\
SN 2008aq & 10 & [4, 5] \\
SN 2008ax & 79 & [4, 8, 9, 10, 11] \\
SN 2011dh & 64 & [5, 12, 13, 14, 15, 16] \\
SN 2011fu & 44 & [5, 13, 17, 18] \\
SN 2012dy & 48 & [13, 19] \\
SN 2013cu &  6 & [13, 20] \\
SN 2013df & 23 & [5, 19, 21, 22] \\
PS15bgt   & 15 & [5, 13] \\
\hline
Total & 432 & \\
\hline
\end{tabular}

\par\smallskip
\begin{minipage}{\columnwidth}
\footnotesize
\raggedright
References:
[1] \citealt{osc2000AJ1201499M};
[2] \citealt{osc1995AAS110513B};
[3] \citealt{osc2015AA573A12J};
[4] \citealt{osc2014AJ14799M};
[5] \citealt{osc2019MNRAS4821545S};
[6] \citealt{osc2001AJ1211648M};
[7] \citealt{osc2009PASP121689S};
[8] \citealt{osc2011MNRAS4132140T};
[9] \citealt{osc2008MNRAS389955P};
[10] \citealt{osc2011ApJ73941C};
[11] \citealt{osc2010ApJ7091343M};
[12] \citealt{osc2014AA562A17E};
[13] \citealt{osc2012PASP124668Y};
[14] \citealt{osc2011ApJ742L18A};
[15] \citealt{osc2015AA580A142E};
[16] \citealt{osc2013MNRAS4363614S};
[17] \citealt{osc2015MNRAS45495M};
[18] \citealt{osc2013MNRAS431308K};
[19] \citealt{osc2016PASA3355C};
[20] \citealt{osc2014Natur509471G};
[21] \citealt{osc2016arXiv160408046S};
[22] \citealt{osc2014AJ14737V}.
\end{minipage}
\end{table}

\begin{figure*}
    \centering
    \includegraphics[width=\linewidth]{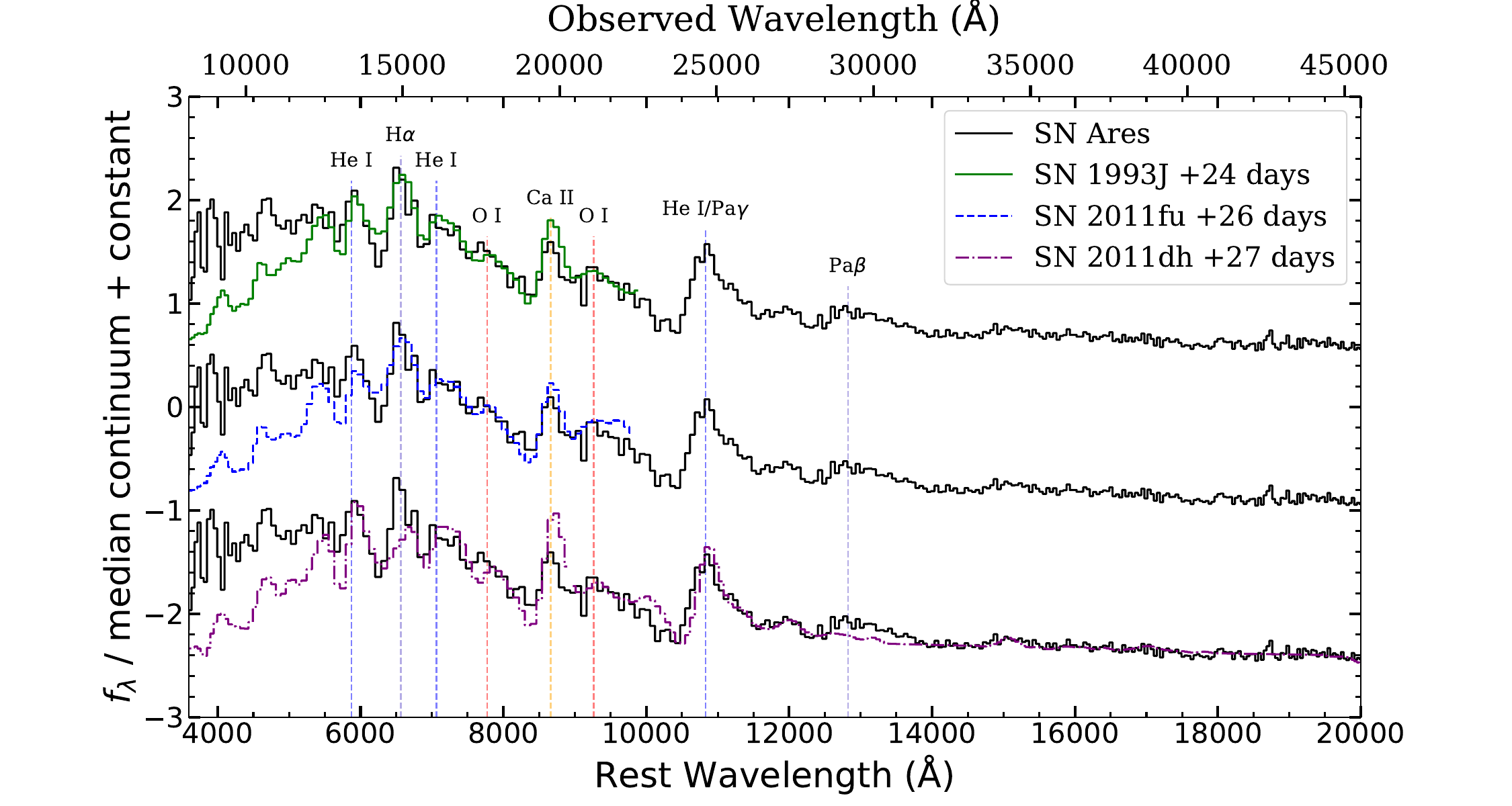}
    \caption{Comparison of well-matching spectra, choices of comparison for which are from an NGSF comparison of \ares\ to the best-observed local SNe~IIb. The SNe are each normalized to their continuua at $7400-7650$ \AA\ rest-frame. The spectrum includes a local background subtraction; residual host contamination is likely the cause of the divergent blue continuum between \ares\ and the comparison SNe. Each SN shows a different ratio between the He I 5876, 7065 \AA\ and H$\alpha$ emission lines. \ares\ appears to have both a slightly stronger H$\alpha$ and weaker Ca II triplet (8498 \AA, 8542 \AA, and 8662 \AA) than the comparison SNe IIb, with the most comparable being SN~1993J. This, along with a seemingly stronger H$\alpha$ profile may point to a less-stripped progenitor hydrogen envelope for \ares, although the strengths of these lines are degenerate with phase. }
    \label{fig:iib_comp}
\end{figure*}

With the classification of \ares as an SN~IIb, we now attempt to find its best match to well-studied, local analogs. To achieve this, we modify NGSF to contain a larger sample of template spectra across a wide phase range. We add all available spectra from the Open Supernova Catalog (OSC) for the SNe~IIb that are included in the most recent NGSF template release \citep{guillochon_open_2017,Goldwasser:2022}. These include: 1993J, 1996cb, 2001ig, 2008aq, 2008ax, 2011dh, 2011fu, 2012dy, 2013cu, 2013df, and PS15bgt (with added spectra and sources shown in Table \ref{tab:osc_iib_spectra}). We fit NGSF in the range of $5000$~\AA\ to $7800$~\AA\ to compare based only on the He I $5876~$\AA, $7065~$\AA, and H$\alpha$ line region, the shape of which strongly correlates with spectral phase. To approximate the resolution of the NIRSpec spectrum, we set the binning of these spectra to 50~\AA; however, we note that such a comparison is coarse and thus we only use this as a qualitative approach to identify possible best analogs to \ares. Our approach yields best matches to the well-studied SN~1993J, SN~2011dh, and SN~2011fu, appearing to be in a phase range of +$15$ - $35$ days post-radioactive peak. We show a few of the closest comparisons, normalized to the median of their continuua between $7400$ to $7650$~\AA, in Figure~\ref{fig:iib_comp}. Overall, the H$\alpha$ line appears stronger for \ares compared to these other SNe, while the Ca II triplet (at $8500$~\AA) appears weaker. Since the SN and host galaxy fill a large fraction of the slit, the sky background was estimated from a small region. We believe imperfect sky subtraction therefore affects flux levels, particularly at the bluest wavelengths where the SN signal is weakest. This flux discrepancy can be seen in the comparison figure, where SN Ares appears to follow the same continuum shape as the comparisons but is overbright on the blue end. Due to this, we do not speculate on differences in any lines blueward of the He I $5876$~\AA\ feature.

\section{Estimate of phase \label{sec:phase}}

For both the purposes of investigating SN physics and utilizing \ares\ as a future time-delay cosmography probe, it is crucial to estimate its phase at the time of our observations. However, unlike SNe Ia, which exhibit a fairly standard light curve evolution, core-collapse SNe display significant heterogeneity in both light curve and spectroscopic evolution, making phase estimation difficult. In Section~\ref{sec:spec_phase}, we estimate the phase of \ares\ for this detected image by comparing against a large sample of stripped-envelope SN spectral templates. We further compare against the best-studied local SN IIb sample and attempt to find our best spectroscopic analogs against which to compare. In Section~\ref{sec:lc_phase}, we use these analogs as light curve templates, as well as templates constructed from the SN IIb population, with which we can more accurately measure the phase.

\subsection{Spectroscopic comparison} \label{sec:spec_phase}

\begin{figure*}
    \centering
    \includegraphics[width=\linewidth]{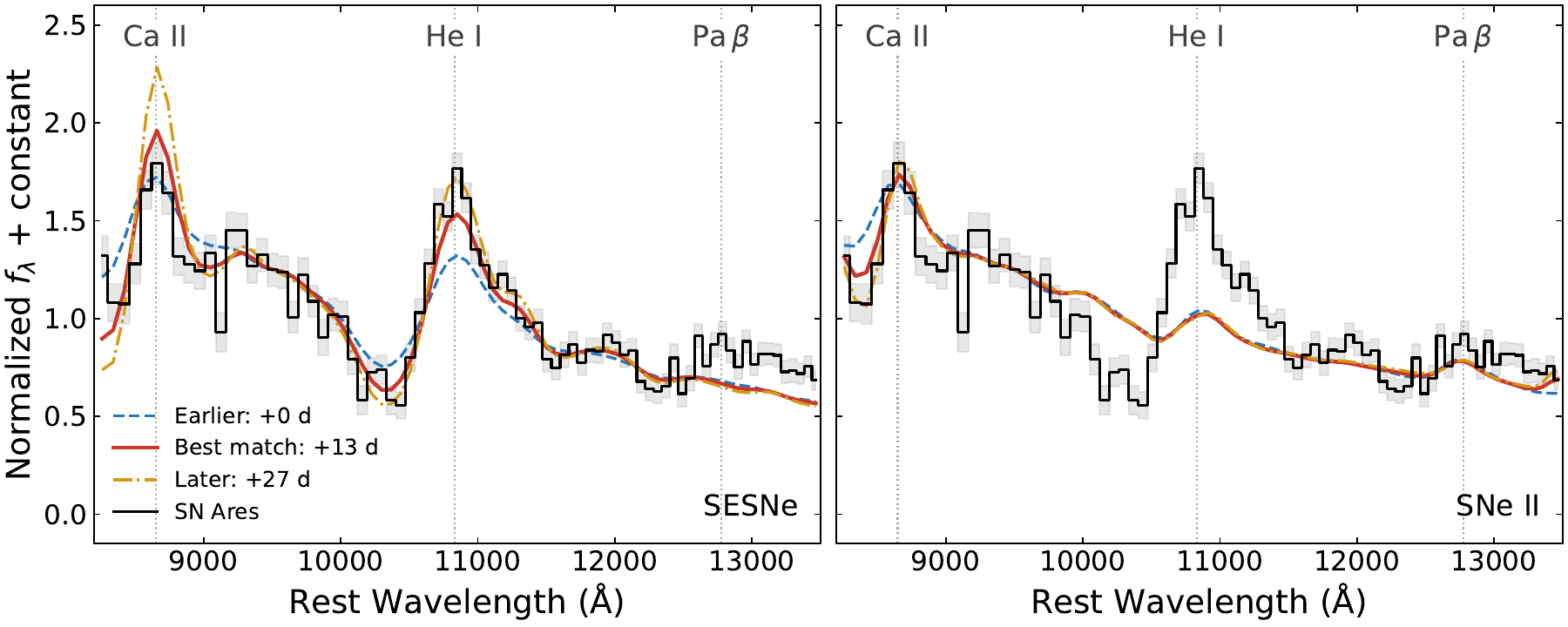}
    \caption{Comparison of SN~Ares versus the NIR templates from the helium-rich CSP SESN sample from \citet{Shahbandeh:2022} on the left and the SN II sample from \citet{davis2019} on the right. \ares\ clearly matches preferentially to the SESN templates, most starkly accentuated by the agreement in the He I feature. Even later time SN~II templates still show strong disagreement with the \ares\ spectrum. The Pa$\beta$ feature does seem to match the SN II relative strength better, possibly indicating that \ares\ retained more of its hydrogen envelope than the low-redshift SESN analogs.}
    \label{fig:temp_comp}
\end{figure*}

Our first approach is to compare against the Carnegie Supernova Project (CSP) stripped-envelope supernova (SESN) templates in the NIR presented by \citet{Shahbandeh:2022}. We specifically utilize the helium-rich templates which were constructed using principal component analysis of both SN IIb and SN Ib NIR spectra, which were shown to be identical up to a fourth order component. These templates are sampled at a 1 day temporal resolution, from -5 days pre-radioactive peak to +80 days post-peak, resulting in 86 templates total. To compare against these templates, we isolate the rest-frame $8200$--$12000$ \AA\ region, encompassing the Ca~II triplet and the He~I/Pa$\gamma$ blend near $10830$ \AA. We use this wavelength region specifically, as the templates show the greatest variation in these two features across the phase range, with the Ca II feature growing in strength relative to the continuum over time, while He I appears to slightly weaken in comparison. To estimate an uncertainty on phase, we calculate a $\chi^2_{\nu}$ value for the difference between the template and \ares\ spectra, after normalizing each to the 9200 -- 9750 \AA\ continuum region. Due to the one-day cadence on the spectral templates, we then simply interpolate through these values to find the phases at which $\chi^2_{\nu} = \chi^2_{\nu,min}+1$ to recover our $1\sigma$ uncertainties. For this, we use a piecewise cubic hermite interpolating polynomial, employed through the \texttt{scipy} package \citep{2020SciPy-NMeth}. We find a spectroscopic phase (relative to \textit{B}-band peak) of $13 \pm 12$ days, with $\chi^2_{\nu,\,\rm min}\approx2.4$. We show the best-matching $+13$-day template along with an earlier and later template from \citet{Shahbandeh:2022} in the left panel of Figure~\ref{fig:temp_comp}. We show the equivalent CSP templates from the SN~II population presented by \citet{davis2019} in the right panel for comparison. From the strong helium emission relative to Pa$\beta$, the classificaton of SN~Ares as an SN IIb is clear. Although the fits to the He-rich SESN templates are overall much better than to the more hydrogen rich SNe II, there are still some differences. For one, the Ca II to He I line ratio differs from the overall population, with the relative strengths of the two lines not perfectly aligning at any template phase. In particular, the Ca II complex strength seems most closely matched at earlier epochs, while the He I feature only begins to match at phases $>20$ days. Thus, fitting to only the He I feature favors a later phase estimate. It is also clear that the Pa$\beta$ line is stronger in \ares\ than the overall SESN population and more similar to the non-stripped SN II population, potentially indicating that \ares\ retains more of its hydrogen envelope than local SESNe on average. We therefore caution that our comparisons against best-matched spectroscopic analogs may be more accurate than this population-level comparison.

As a further comparison, we turn to the best local analogs, SN~1993J, SN~2011dh, and SN~2011fu. We focus on the optical He I and H$\alpha$ line complex, since the relative strengths of these lines also vary by phase. For this analysis, we first fit and subtract the continuum around the region and then normalize to the median of the line region, which we define as being from $5600 - 7500$~\AA. We use all available spectra from the OSC (references shown in Table~\ref{tab:osc_iib_spectra}) and fit for the $\chi^2_{\nu}$ as with the CSP comparison. We require that the comparison spectra cover the full wavelength range of the line complex. We then estimate the best phase by identifying the minimum $\chi^2_{\nu}$ value through a polynomial fit between the phases and $\chi^2_{\nu}$ values (after removing $\chi^2_{\nu}$ values above 10) using \texttt{numpy} and again define our $1\sigma$ uncertainties as based on where $\chi^2_{\nu} = \chi^2_{\nu,\,\rm min}+1$. From this, we find phase estimates of $30\pm19$, $23 \pm 20$, and $15 \pm 26$ days for the three SNe, respectively. Taking the weighted average of these values, we obtain a phase estimate of $24\pm12$ days, a wide range but in agreement with the NIR CSP comparison. In particular, this analysis favors a later phase estimate, such as we obtain in the NIR when we only fit the He I feature. We find that SN~1993J appears to be the closest analog to \ares\ of the three, with the lowest $\chi^2_{\nu,\,\rm min}$ minimum of 1.9.

\subsection{Light curve fits} \label{sec:lc_phase}

To estimate the phase of SN Ares photometrically, we apply a similar methodology as for the spectroscopic comparison: fitting against both population-level templates and the best local analog SNe IIb. For the latter, we compare the \ares\ photometry to the previously discussed SNe IIb 1993J, 2011dh, and 2011fu. The comparison light curves are corrected for Milky Way and host-galaxy extinction before fitting. We adopt total color excesses of $E(B-V)=0.194$ mag for SN 1993J \citep{Lyman2014}, $E(B-V)=0.07$ mag for SN 2011dh \citep{Ergon_2014}, and $E(B-V)=0.10$ mag for SN 2011fu \citep{Morales-Garoffolo_2015}. The SN Ares photometry are corrected for a fixed Milky Way foreground extinction of $E(B-V)_{\rm MW}^{}=0.184$ mag \citep{schlegel_new_1990,schlafly_measuring_2011} assuming $R_V=3.1$ and a Fitzpatrick extinction law \citep{fitzpatrick_correcting_1999}. Any additional extinction along the line of sight to Ares is added as a free parameter in our light curve fitting methodology. The sources of the photometry for each comparison SN were obtained by searching the OSC, with individual sources and bands described in Table~\ref{tab:ares_photometry_sources}.

For each comparison SN, we construct Gaussian-process (GP) interpolations of the available light curves. The GP model uses a Mat\'ern-$3/2$ covariance kernel and includes the reported photometric uncertainties from the OSC. The time of the main radioactive $B$-band maximum is determined from the GP interpolation of the $B$-band light curve around the radioactive peak.

\begin{deluxetable}{lcc}
\tabletypesize{\small}
\tablewidth{\columnwidth}
\tabcolsep=5pt
\tablecaption{
Photometric templates used for SN~Ares
\label{tab:ares_photometry_sources}
}
\tablehead{
\colhead{Comparison} &
\colhead{Template bands} &
\colhead{Refs.}
}
\startdata
SN~1993J       & $V,R,I,K$ & 1--8 \\[7pt]
SN~2011dh      & $g,r,z,K$ & 9    \\[7pt]
SN~2011fu      & $V,R,I$   & 10   \\[7pt]
IIb population & $g,R,I$   & 11   \\
\enddata

\tablecomments{
All comparisons use the \textit{JWST} filters F115W, F150W, and F200W, mapped in order to the first three template bands. The SN~1993J and SN~2011dh comparisons additionally use F444W, mapped to $K$. The $B$-band templates are used separately to define phase zero. For SN~1993J, references 1--7 provide optical photometry and reference 8 provides the provisional $K$-band attribution.
}

\tablerefs{
(1) \citet{Richmond1994};
(2) \citet{Richmond1996};
(3) \citet{Barbon1995};
(4) \citet{Benson1994};
(5) \citet{vanDriel1993};
(6) \citet{Metlova1995};
(7) \citet{Okyudo1993};
(8) \citet{Matthews2002};
(9) \citet{Ergon2014};
(10) \citet{Kumar2013};
(11) \citet{Khakpash2024}.
}


\end{deluxetable}

\begin{figure*}
    \centering
    \includegraphics[width=0.48\linewidth]{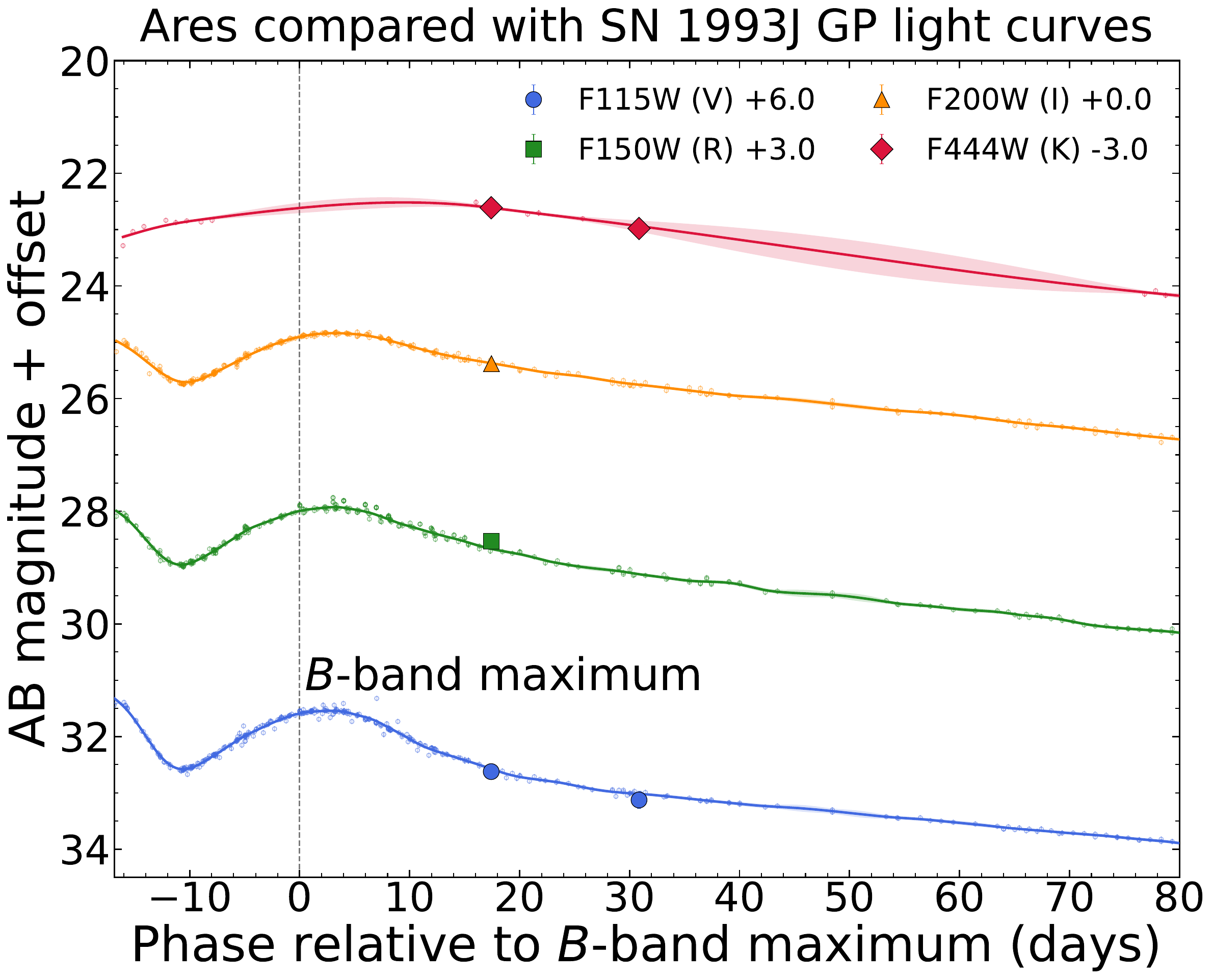}
    \includegraphics[width=0.4825\linewidth]{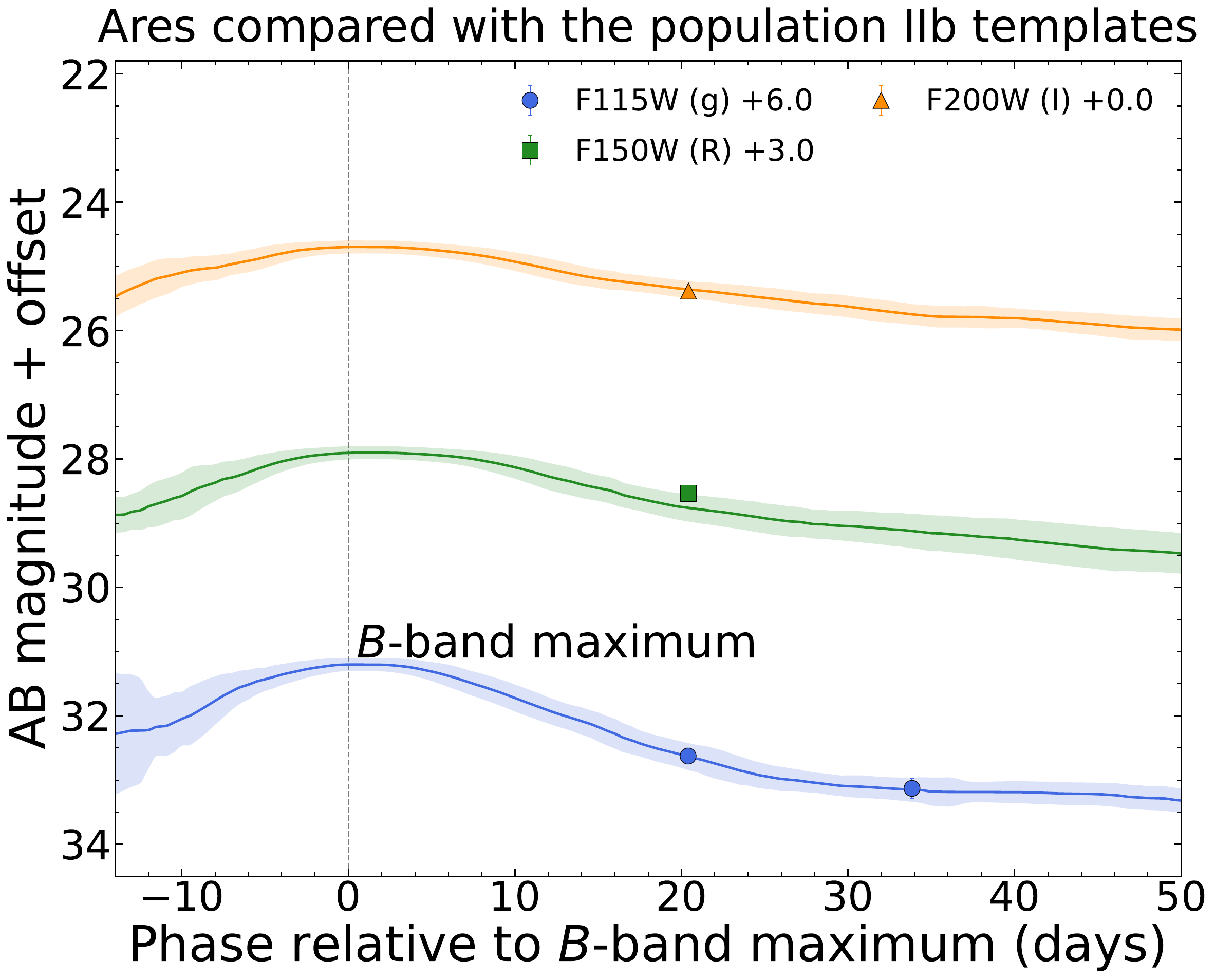}
\caption{Light curve fits against SN~1993J, which showed the best match to \ares\ both spectroscopically and photometrically, and the SN~IIb population-level light curves. We show the \ares\ photometry overlaid on the GP fits at the best estimate of phase. Filter matching choices are shown in the legends in the upper right. All photometry from the OSC for SN~1993J (shown as the smaller points) was corrected for known dust extinction before fitting, while the published GP templates from \citet{Khakpash2024} were used for the population-level comparison. Corner plots for the fit are shown in Appendix~\ref{app:additional_lc_fits}.}
    \label{fig:lc_comp}
\end{figure*}

Because the observed and rest-frame filters are not identical, cross-filter corrections are calculated directly from the observed spectrum of SN Ares. Before calculating these corrections, the spectrum is corrected for Milky Way extinction. Synthetic photometry through the appropriate observer- and rest-frame filter transmission curves is then used to determine the cross-filter correction for each filter pair. We include an additional uncertainty of $0.10$ mag in quadrature to account for systematic uncertainty associated with deriving the cross-filter corrections from a single spectrum, which is based on the largest cross-filter correction value we obtain in our methodology.

For a photometric observation at MJD $t_i$, the rest-frame phase in the comparison-SN light curve is
\begin{equation}
    \phi_i =
    \phi_{\rm ref} +
    \frac{t_i-t_{\rm ref}}{1+z_{\rm Ares}},
\end{equation}
where $z_{\rm Ares}^{}=1.276$ and $t_{\rm ref}$ is the date of the spectrum observation. Thus, the fitted phase reported throughout the analysis corresponds specifically to the phase of SN Ares at the second \textit{JWST} epoch.

The model contains four free parameters: the reference phase $\phi_{\rm ref}$, a common magnitude offset $\Delta m$ that combines both the effect of magnification and the redshift-dependent \textit{K} correction, the host-galaxy color excess of SN Ares $E(B-V)_{\rm host}$, and an additional intrinsic-scatter term $\sigma_{\rm int}$. For an observation in filter $f$, the model magnitude is schematically
\begin{equation}
    m_{{\rm model},i}
    =
    m_{{\rm GP},f}^{}(\phi_i)
    + \Delta m
    + K_f
    + A_f,
\end{equation}
where $m_{{\rm GP},f}^{}$ is the GP-interpolated comparison-SN magnitude, $K_f$ is the spectrum-derived cross-filter correction, and $A_f$ is the extinction in the corresponding rest-frame band.

The likelihood is evaluated using
\begin{equation}
    \ln{\cal L}
    =
    -\frac{1}{2}
    \sum_i
    \left[\ln(2\pi \sigma_i^2)+
    \frac{(m_i-m_{{\rm model},i})^2}{\sigma_i^2}
    \right],
\end{equation}
with
\begin{equation}
    \sigma_i^2 =
    \sigma_{m,i}^2
    + \sigma_{{\rm GP},i}^2
    + \sigma_K^2
    + \sigma_{\rm int}^2,
\end{equation}
where $\sigma_K^{}=0.10$ mag.

For SN 1993J, the light curve contains an extended late-time decline that can produce a secondary, weakly constrained phase solution. To prevent this solution from generating an artificial long tail in the phase posterior, we restrict the phase at the second \textit{JWST} epoch to $-10 < \phi_{B,\,\max} < 60$ rest-frame days relative to the radioactive $B$-band maximum. We additionally impose a weak half-normal regularization on the intrinsic-scatter parameter, with characteristic scale $0.15$ mag and $\sigma_{\rm int}<0.35$ mag. Both of these phase restrictions are implemented to prevent poor late-time matches from being accommodated simply by an inflated intrinsic scatter or from very uncertain GP fits. These choices on the prior are also informed by the spectroscopic phase estimates previously described, which strongly disfavor a very early or late phase.

Posterior distributions are sampled using the affine-invariant MCMC sampler implemented in the \texttt{emcee}  package \citep{foreman-mackey_emcee_2013}. We characterize each fitted parameter by the posterior median and report uncertainties based on the 68th percentile confidence intervals. We find that the best-fit individual comparison is to SN~1993J, with a $\chi^2_{\nu}$ close to unity.

For our population-level comparison, we rely on the SN IIb GP templates provided by \citet{Khakpash2024}. These templates were constructed using the total sample of SNe IIb provided by the OSC and is therefore not independent of our previous analyses. Unfortunately, due to lack of data, these models do not include a \textit{K}-band template. Thus, we only fit with the provided matching filters shown in Table~\ref{tab:ares_photometry_sources} for our final phase estimate. We test the \textit{H}-band template to check the decline rate consistency between our observations and the template, and find good agreement at the final estimated phase from the optical-only fit. We show the best-fit single comparison object, SN~1993J, and the population-level template fit in Figure~\ref{fig:lc_comp}.

From these template fits to SNe 1993J, 2011dh, 2011fu and the population-level templates, we recover a phase estimate for the second epoch of \textit{JWST} data (e.g. our spectral phase) of $31_{-5}^{+11}$, $34_{-5}^{+10}$, $31_{-9}^{+13}$, and $34_{-9}^{+10}$ days, respectively. Therefore we find strong consistency in our measured phase across our best three local analogs, as well as from the population as a whole. The posterior distributions (shown in Appendix~\ref{app:additional_lc_fits}) overall seem to show asymmetric distributions on phase, often with a higher-phase mode. This skewed uncertainty is represented in our reported phases, and we note that all of our phase estimates are in $1\sigma$ agreement to our estimates from the spectral comparisons.

\subsubsection{Peak $B$-band magnitude of SN Ares}

We additionally use the fitted magnitude offsets to estimate the intrinsic peak rest-frame $B$-band absolute magnitude of SN Ares. Because the fitted magnitude offset is defined between the apparent light curves of SN Ares and each comparison SN, the intrinsic peak absolute magnitude of SN Ares is calculated as
\begin{equation}
\begin{split}
M_{B,\,\max}^{\rm Ares}
&= m_{B,\,\max}^{\rm tpl,0} + \Delta m
   - \mathrm{DM}_{\rm Ares} \\
&\quad + 2.5\log_{10}\mu
   + 2.5\log_{10}(1+z).
\end{split}
\label{eq:ares_absolute_B}
\end{equation}

Here, $M_{B,\,\max}^{\rm Ares}$ is the intrinsic, unlensed absolute magnitude of SN Ares at rest-frame $B$-band maximum, and $m_{B,\max}^{\rm tpl,0}$ is the extinction-corrected apparent $B$-band peak magnitude of the comparison template. The parameter $\Delta m$ is the common magnitude offset obtained by fitting the template light curves to Ares, with positive values corresponding to a fainter source. The distance modulus is $\mathrm{DM}_{\rm Ares}=5\log_{10}(D_L/10\,\mathrm{pc})$, where $D_L$ is the luminosity distance in the adopted cosmology, which at $z=1.276$ gives a distance modulus of approximately 44.76~mag. The magnification $\mu$, which we take to be $5.6\pm0.5$ from the \texttt{Lenstool} model, enters through $+2.5\log_{10}\mu$ and corresponds to a $\sim1.9$ mag difference. Finally, $+2.5\log_{10}(1+z)$ corrects for the redshift-dependent \textit{K}-correction at the \ares\ redshift. The Ares Milky Way extinction and cross-filter corrections have already been applied to the photometry, while the \ares\ host extinction is included explicitly in the light-curve model.

We propagate the uncertainty in the fitted magnitude offset together with the asymmetric uncertainty in the lensing magnification using Monte Carlo sampling, with uncertainties propagated to the $M_{B,\,\max}$ absolute magnitude estimates. In the Vega system (with AB, Vega offsets from \citealt{Blanton2007}) these estimates for SN~1993J, SN 2011dh, and SN 2011fu are $-17.9_{-0.3}^{+0.2}$, $-17.1_{-0.3}^{+0.2}$, and $-17.4_{-0.4}^{+0.4}$, respectively. These values are all well within the distribution of SN IIb peak magnitudes presented by \citet{richardson_absolute_2014}, where $\langle \rm{M}_B\rangle=17.0\pm1.3$ mag. Using the magnification from the \texttt{glafic} model produces slightly brighter estimates, but still within expectations.

\section{Discussion \& Conclusions}
\label{sec:discussion}

Our analysis of SN Ares demonstrates the success of targeted \textit{JWST} observations of rich galaxy clusters for the discovery and detailed study of high-redshift SNe. At a spectroscopic redshift of $z=1.276$, Ares provides an unprecedented view of a high-redshift SN IIb through simultaneous coverage of its rest-frame optical and near-infrared spectrum. The combination of this wavelength coverage, multiband photometry, and a multiply imaged host galaxy makes the system valuable both for investigating SN properties at high redshift and for high-precision time-delay cosmography.

The classification of \ares\ as an SN IIb is supported by the presence of hydrogen and helium features and their relative strengths. In particular, the optical He~I features near $5876$ and $7065$~\AA, together with the near-infrared He~I feature near $10830$ \AA, favor a helium-rich explosion that retains some hydrogen. Although these individual features also appear in regular SNe II at later times, their collective morphology and relative prominence are best-matched to an SN IIb classification. The absence of prominent forbidden lines expected from a substantially later-phase SN II further supports this interpretation.

In order to estimate the phase, we compare SN Ares to three local analogs: SN 1993J, SN 2011dh, and SN 2011fu, all chosen based on their spectroscopic similarities. Among the local comparison objects, SN~1993J provides the closest overall match, although SN Ares is still distinct. Compared to these analogs, \ares\ seems to possess stronger H$\alpha$ emission and a weaker Ca~II near-infrared triplet, which could reflect differences in envelope structure, mass, and excitation, although these properties all also depend on phase. Residual host-galaxy emission may contribute to the observed H$\alpha$ profile, making differences in hydrogen envelope mass poorly constrained given the current dataset. A future \textit{JWST}/NIRSpec MOS observation is scheduled for early 2027 (PID 4568), which will allow for a host spectrum to be obtained after the transient has faded. This will allow us to subtract the host spectrum directly at the location of SN Ares and will help distinguish intrinsic SN emission from contamination, enabling a more reliable comparison of the line profiles. The photometric comparisons consistently place the NIRSpec observation approximately one month after the \textit{B}-band radioactive peak. Fits using SN~1993J, SN~2011dh, and SN~2011fu give median phases between $31$ and $34$ rest-frame days, consistent with a population-template estimate of $34^{+10}_{-9}$ days using published GP templates constructed from all SN IIb light curves from the OSC. Our spectroscopic estimates of phase are much broader but still support a late-photospheric interpretation.

After correcting for the model magnification, the inferred peak $B$-band magnitudes span approximately $-17.1$ to $-17.9$~mag across the three individual comparisons. These estimates are consistent with the luminosity distribution of nearby SNe IIb and do not require an unusually luminous explosion. The magnification of approximately $\mu=5.6$, corresponds to a brightness enhancement of about $1.9$~mag, which enables spectroscopic study of an event whose inferred luminosity lies within the local SN IIb distribution. Its broad resemblance to nearby analogs shows that familiar SN IIb spectral behavior can be identified at this redshift, although a single event cannot establish whether the population or its progenitor channels have evolved. Apart from any individual physical differences to local analogs; however, it is perhaps interesting to find an SN IIb among the limited ($<10$) spectroscopically observed strongly lensed SN sample, given that rates of SNe IIb are relatively low in the local Universe \citep[$\sim7\%$ of SNe II,][]{Gonzalez-Banuelos_iib_2025}. Given its similarity to SN 1993J, 2011dh, and 2011fu, which have all been shown to have binary progenitor systems, it is likely that SN Ares has a similar higher-multiplicity progenitor origin. SN Refsdal, a 1987A-like SN, arose from an even rarer progenitor system that also almost certainly had a multiplicity greater than one. Future spectroscopic observations of lensed SNe may show whether such a trend in higher multiplicity at higher redshfits exists, which may be expected given the predicted inverse relationship between multiplicity and metallicity. More spectroscopic observations of lensed SNe are needed to disentangle these rates, as the magnification is needed to detect such objects at these redshifts. For blank-field surveys without the benefit of magnification, SNe like Refsdal and Ares would likely have gone undetected or with spectrophotometric datasets that were insufficient for a correct classification.

In addition to the interesting physical nature of SN Ares, its potential use as a long-baseline time-delay cosmography probe is also of particular note. Our current best lens models predict future appearances of SN Ares in $\sim73$ and $76$ years after the observed image. Additional spectroscopic redshifts for the multiply imaged systems which will be obtained in early 2027 will strengthen the lens constraints and help assess the stability of the predicted delays. Assuming that the delays remain on the order of 70 years, the SN Ares system has by far the highest predicted delays of any discovered lensed SN system so far. The reappearance of SN Ares, and other yet-to-be discovered long-baseline lensed SNe, will require dedicated monitoring observations from next-generation telescopes. For a telescope set to launch within the next few decades, such as the \textit{Habitable Worlds Observatory}, guaranteeing the reappearance of \ares\ may require the capability to service the mission to extend its lifetime. Assuming we are able to catch its reappearance, it would be trivial to measure a negligible, $\ll1\%$ uncertainty on the time delay for \ares\ given its planned final dataset. Since the uncertainty on the time delay is directly proportional to the uncertainty on measured cosmological parameters, SN Ares represents the possibility for the most precise single-step measurement of cosmology currently known. The usefulness of SN Ares for cosmology will be limited only by the systematics in the lens models themselves. Our understanding of cosmology and lens modeling in 70 years is impossible to predict but regardless, the reappearance of SN Ares will almost certainly be a scientifically rich event worth capturing.

\begin{acknowledgments}

CL acknowledges support from HST-GO-17474 and JWST-DDT-9478.

This work was supported by JSPS KAKENHI Grant Numbers JP25H00662, JP22K21349.

MB acknowledges support from the ERC Grant FIRSTLIGHT \# 101053208, Slovenian national research agency ARIS through grants N1-0238 and P1-0188, and ESA PRODEX Experiment Arrangements No. 4000146646 and 4000149972.

P. Dayal warmly acknowledges support from an NSERC discovery grant (RGPIN-2025-06182).

SF acknowledges support from the Dunlap Institute, funded through an endowment established by the David Dunlap family and the University of Toronto. We acknowledge the support of the Canadian Space Agency (CSA) [25JWGO4A18].

TK acknowledges support by KAKENHI (26KJ1232) through Japan Society for the Promotion of Science (JSPS).

MN acknowledges support from KAKENHI Grant Nos. 25KJ0828 through Japan Society for the Promotion of Science (JSPS).

MO acknowledges the supports from the World Premier International Research Center Initiative (WPI Initiative), MEXT, Japan, the joint research program of the Institute for Cosmic Ray Research (ICRR), the University of Tokyo, and KAKENHI (20H00180, 21H04467, 25H00674) through Japan Society for the Promotion of Science (JSPS).

HY acknowledges support by KAKENHI (25KJ0832) through Japan Society for the Promotion of Science (JSPS).

FEB acknowledges support from ANID-Chile BASAL CATA FB210003 and FONDECYT Regular 1241005.

CJC acknowledges support from the ERC Advanced Investigator Grant EPOCHS (788113).

J.M.D. acknowledges support from project PID2022-138896NB-C51 (MCIU/AEI/MINECO/FEDER, UE).

YF acknowledges supports from JSPS KAKENHI Grant Numbers JP23K13149 and 26K00736.

MGO acknowledges financial support from the State Agency for Research of the Spanish MCIU through Center of Excellence Severo Ochoa award to the Instituto de Astrofísica de Andalucía CEX2021-001131-S funded by MCIN/AEI/10.13039/501100011033, and from the grant PID2022-136598NB-C32 ``Estallidos8'' funded by MCIN/AEI/10.13039/501100011033 and by ``ERDF A way of making Europe''. MGO also acknowledges the support by the project ref. AST22\_00001\_Subp\_11 funded from the EU -- NextGenerationEU, PPCC Junta de Andalucía.

PAAL thanks the support from CNPq, grants 310260/2025-6 and 404160/2025-5, and FAPERJ, grant E-26/200.545/2023.

GEM acknowledges support from the Villum Fonden research grants 37440 and 13160 and by the Danish National Research Foundation under grant DNRF140 (Cosmic DAWN Center).

MM acknowledges financial support through grants INAF GO Grant 2022 ``The revolution is around the corner: JWST will probe globular cluster precursors and Population III stellar clusters at cosmic dawn'' and by the European Union -- NextGenerationEU within PRIN 2022 project n.20229YBSAN - Globular clusters in cosmological simulations and lensed fields: from their birth to the present epoch.

RPN is grateful for the generous support of Neil and Jane Pappalardo through the MIT Pappalardo Fellowship.

RAW acknowledges support from NASA JWST Interdisciplinary Scientist grants NAG5-12460, NNX14AN10G and 80NSSC18K0200 from GSFC.

YZ acknowledges support from the MAOF grant (no. 12641898) and visitor support from the Observatories of the Carnegie Institution for Science, Pasadena, CA, where part of this work was completed.

AZ acknowledges support by the Israel Science Foundation Grant No. 864/23.

RA acknowledges support of Grant PID2023-147386NB-I00 funded by MICIU/AEI/10.13039/501100011033 and by ERDF/EU, and the Severo Ochoa award to the IAA-CSIC CEX2021-001131-S and from grant PID2022-136598NB-C32 ``Estallidos8''.

GK acknowledges computing support by the University of Pittsburgh Center for Research Computing, RRID:SCR 022735, through the resources provided; specifically, this work used the H2P/MPI cluster, which is supported by NSF award number OAC-2117681.

Y.J-T. acknowledges financial support from the State Agency for Research of the Spanish MCIU through Center of Excellence Severo Ochoa award to the Instituto de Astrofísica de Andalucía CEX2021-001131-S funded by MCIN/AEI/10.13039/501100011033, and from the grant PID2022-136598NB-C32 Estallidos and project ref. AST22-00001-Subp-15 funded by the EU-NextGenerationEU.

\end{acknowledgments}




%
\facilities{JWST (NIRSpec and NIRCam),  HST (WFC3 UVIS and IR)}

\software{
Astropy \citep{astropycollaboration_astropy:_2013,
astropycollaboration_astropy_2018, AstropyCollaboration2022},
Astroquery \citep{ginsburg_astroquery_2019},
Matplotlib \citep{hunter_matplotlib:_2007},
NumPy \citep{oliphant_guide_2006},
SciPy \citep{virtanen_scipy_2020},
pandas \citep{mckinney_pandas_2010},
scikit-learn \citep{pedregosa_scikit_learn_2011},
emcee \citep{foreman_mackey_emcee_2013},
corner.py \citep{foreman_mackey_corner_2016},
dust\_extinction \citep{karl_gordon_2022_6397654},
extinction (\url{https://github.com/kbarbary/extinction}),
Photutils \citep{photutils},
space\_phot (\url{https://github.com/justincpierel/space_phot}),
jwst \citep{Bushouse_JWST_Calibration_Pipeline_2024},
grizli \citep{brammer2023},
msaexp \citep{Brammer23},
NGSF \citep{Goldwasser:2022},
Lenstool \citep{jullo_lenstool_2007},
glafic \citep{oguri2010,oguri2021b},
AstroLensPy \citep{allingham26},
ChatGPT \citep{openai_chatgpt_2026}
}


\clearpage

\appendix

\section{HST Analysis}
\label{app:hst_upper}

\begin{figure*}[!tp]
    \centering
    \includegraphics[width=0.8\linewidth]{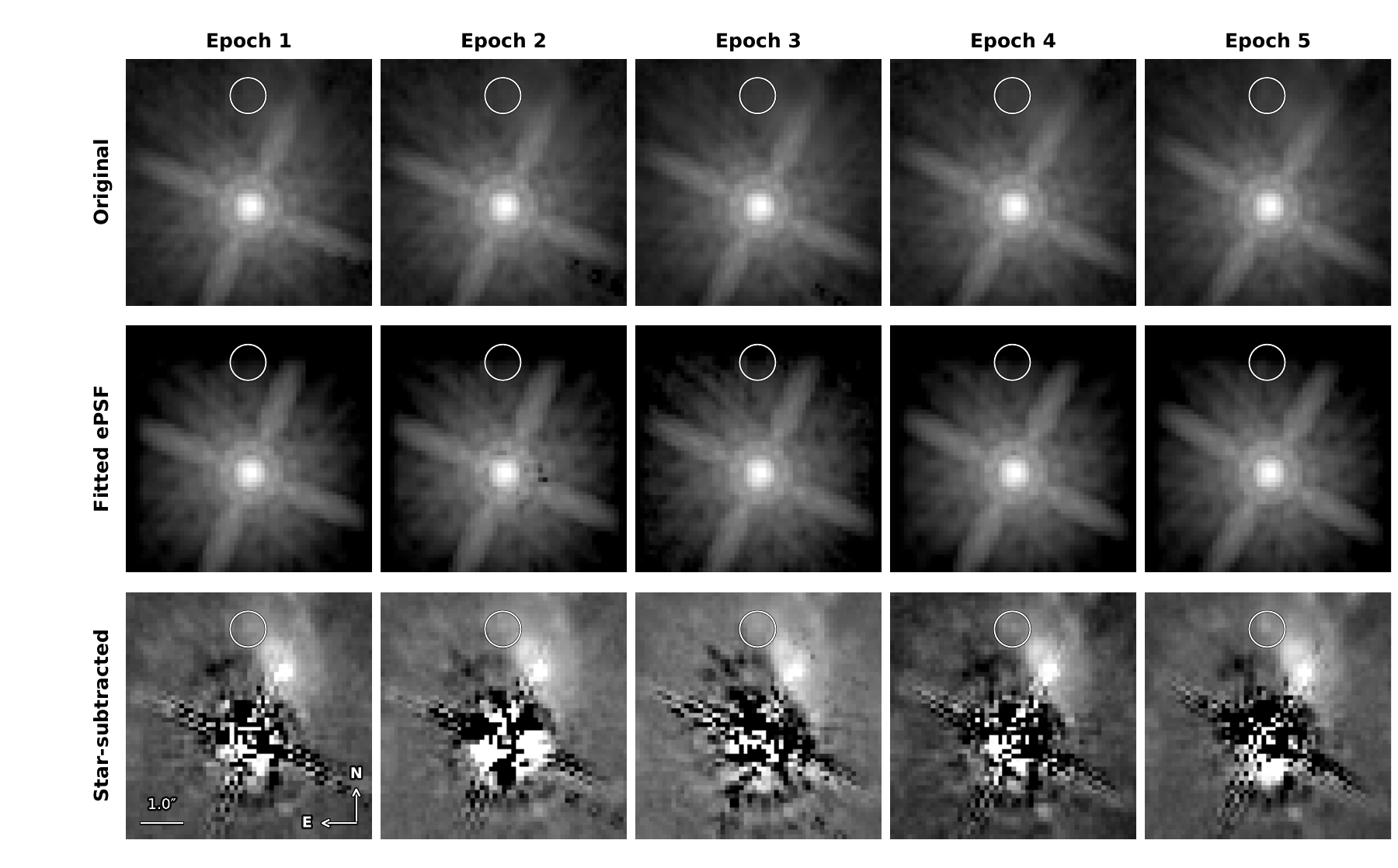}
    \caption{PSF subtraction process for the five science epochs in F160W. The stark difference in the appearance of the host galaxy shows how strong the stellar contamintation is in the \textit{HST} data. The predicted location of SN Ares from the \textit{JWST} data is marked as a circle.}
    \label{fig:psf_sub}
\end{figure*}

As mentioned in Section~\ref{sec:obs}, five \textit{HST}/WFC3 F814W, F110W, and F160W observations of \ares\ were obtained following its discovery. Unfortunately, due to the bright nearby foreground star, these observations result in nondetections. We constrain the F160W flux of SN Ares using the five \textit{HST} follow-up epochs and a pre-explosion template from the Reionization Lensing Cluster Survey \citep[RELICS,][]{coe_relics_2019}. We subtract a fitted empirical PSF model for the nearby bright foreground star from both the science and template images using \texttt{space\_phot} \citep{pierel_space-phot_2024}. The star-model amplitudes and centroids are refit while allowing for a varying local background. We correct the pedestal in the saved PSF models and apply a raised-cosine taper over their outermost three pixels to suppress discontinuities at the model boundaries. This process for the science image is shown in Figure~\ref{fig:psf_sub}.

\begin{figure*}[!tp]
    \centering
    \includegraphics[width=0.8\linewidth]{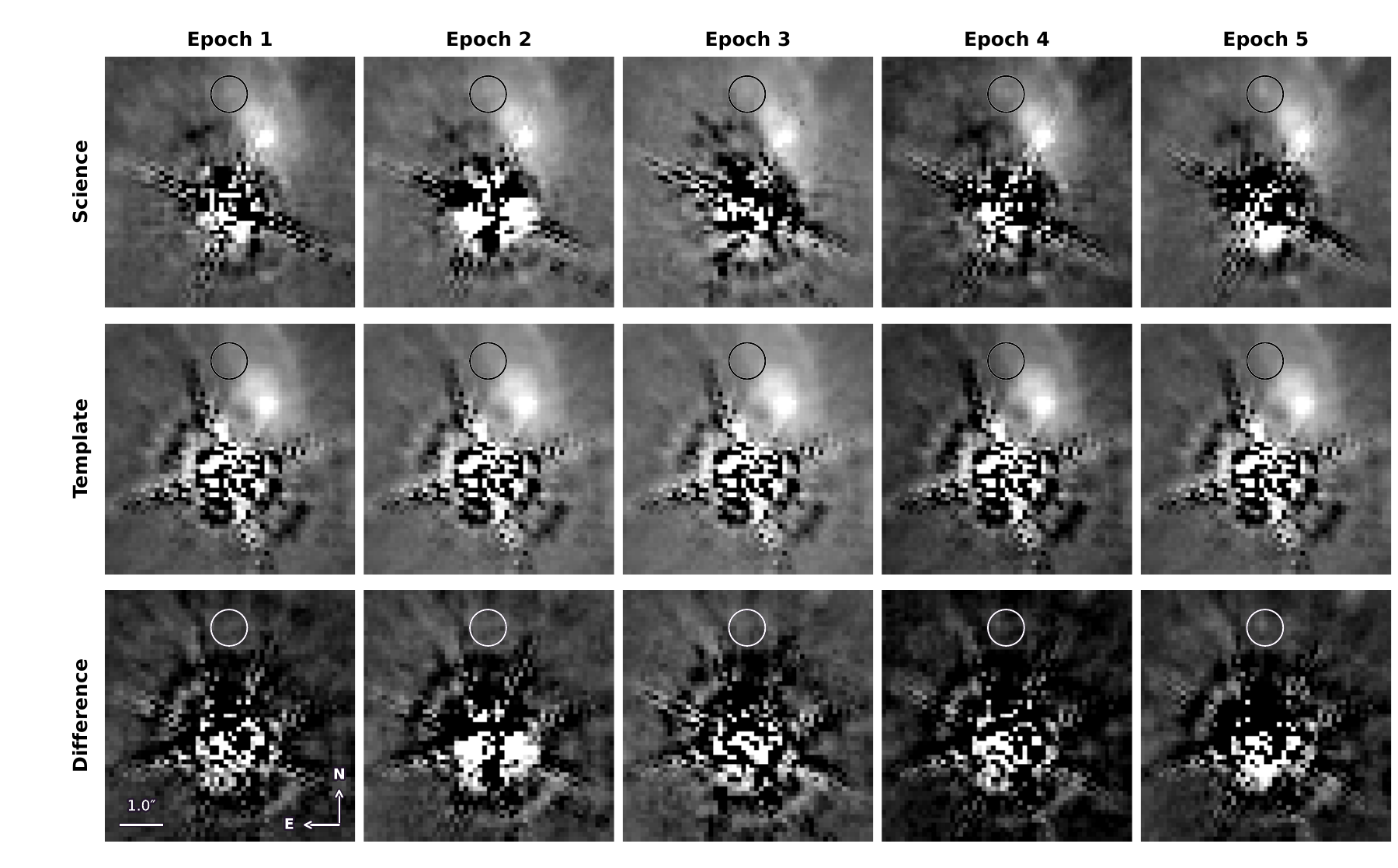}
    \caption{Difference imaging process between the science epochs and template image in F160W. The template image bright star was also subtracted before the final difference image was made. Residuals from both of the images contribute to the noise of the final image, thus making the detection upper limit much less constraining than if a similar-PA template was used.
    }
    \label{fig:hst_sub}
\end{figure*}

The star-subtracted science and template images share a common pixel grid and flux calibration due to the VENUS reduction pipeline. We form direct differences without additional resampling or PSF-matching convolution. We attempted such a deconvolution method as well, which erturned similar results. We measure the flux at the fixed SN Ares position using forced PSF photometry with the empirical science-image PSF and fit within a five-pixel radius. The difference images after PSF subtraction are shown in Figure~\ref{fig:hst_sub}.

We propagate the science and template error maps in quadrature and empirically rescale the formal flux uncertainties using the robust scatter of forced measurements at 131 nearby control positions. We assess sensitivity to the adopted SN position, fitting radius, star-model pedestal correction, model-edge taper, and relative photometric scaling and alignment. The largest absolute change in the fitted flux across these checks is adopted as a systematic term and added in quadrature to the empirically rescaled statistical uncertainty. No epoch exceeds a nominal significance of $3\sigma$. We present the related upper limits in Table~\ref{tab:ares_hst_limits}. We attempt this same approach for the other filters and similarly find nondetections. Because SN Ares is expected to be much fainter in these bands than F160W, we do not include the upper limits, as they are uninformative. Wihtout contamination from the bright, nearby star, it is almost certain that SN Ares would have been detected in at least the first epoch of F160W, given its predicted brightness. A future template image that matches the PA of the science images could be used to produce a much cleaner difference image where possible detections could be made.

\begin{table}[htbp]
\centering
\caption{Approximate $3\sigma$ F160W upper limits for SN Ares from direct subtraction of the star-subtracted \textit{HST} images. Magnitudes are observed AB values, without extinction or lensing corrections.}
\label{tab:ares_hst_limits}
\begin{tabular}{cccc}
\hline
Epoch & MJD & AB magnitude limit \\
 & & (mag) \\
\hline
1 & 60940.48471 & $>25.30$ \\
2 & 60945.52527 & $>25.42$ \\
3 & 60951.61157 & $>25.39$ \\
4 & 60956.71353 & $>25.26$ \\
5 & 60962.40768 & $>24.83$ \\
\hline
\end{tabular}
\end{table}

\clearpage

\section{Redshift Estimation}
\label{app:redshift}

\begin{figure*}[!h]
    \centering
    \includegraphics[width=\linewidth]{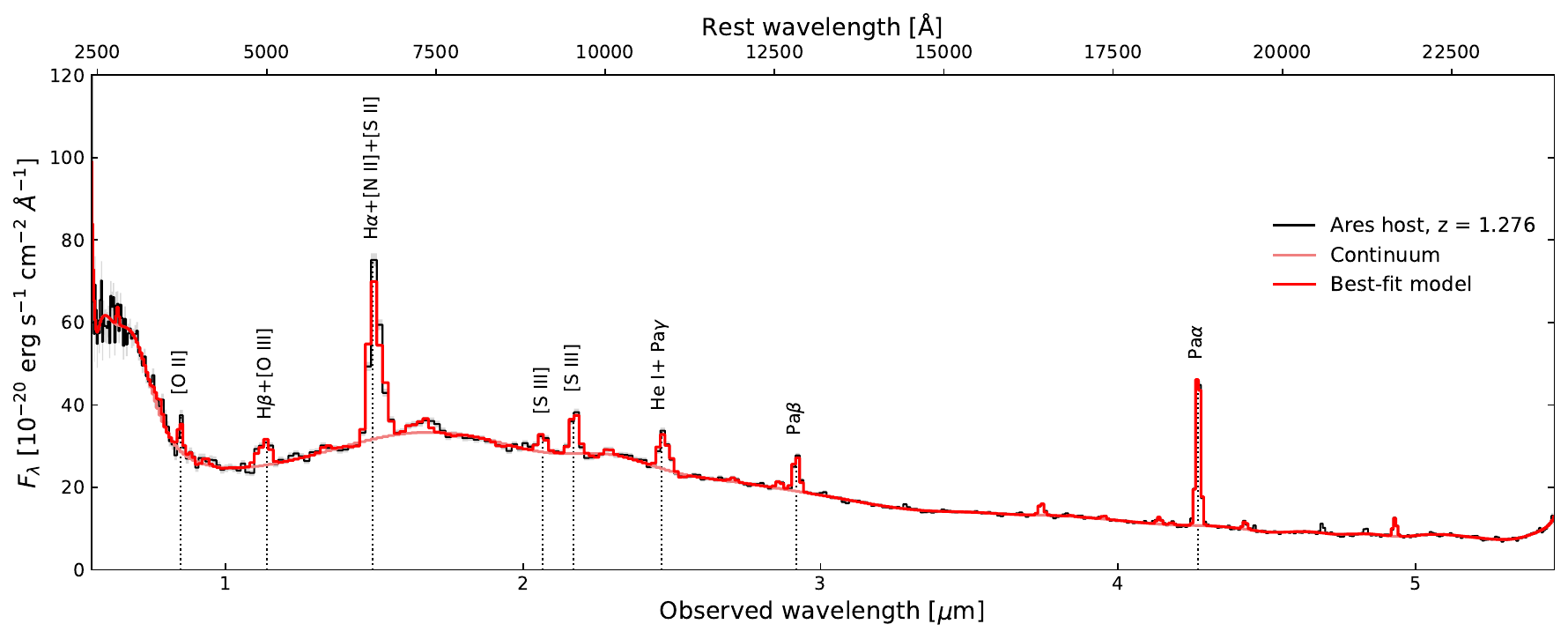}
    \caption{
    Extracted host spectrum with emission lines that were used for redshift estimation overplotted.
    }
    \label{fig:redshift}
\end{figure*}

In Figure~\ref{fig:redshift}, we show the extracted spectrum of the host of \ares\ that was used to obtain our redshift estimate of $z=1.276\pm0.002$

\clearpage

\section{Additional Light-Curve Fits and Posterior Distributions}
\label{app:additional_lc_fits}

\begin{figure*}[!h]
    \centering
    \includegraphics[width=0.49\linewidth]
        {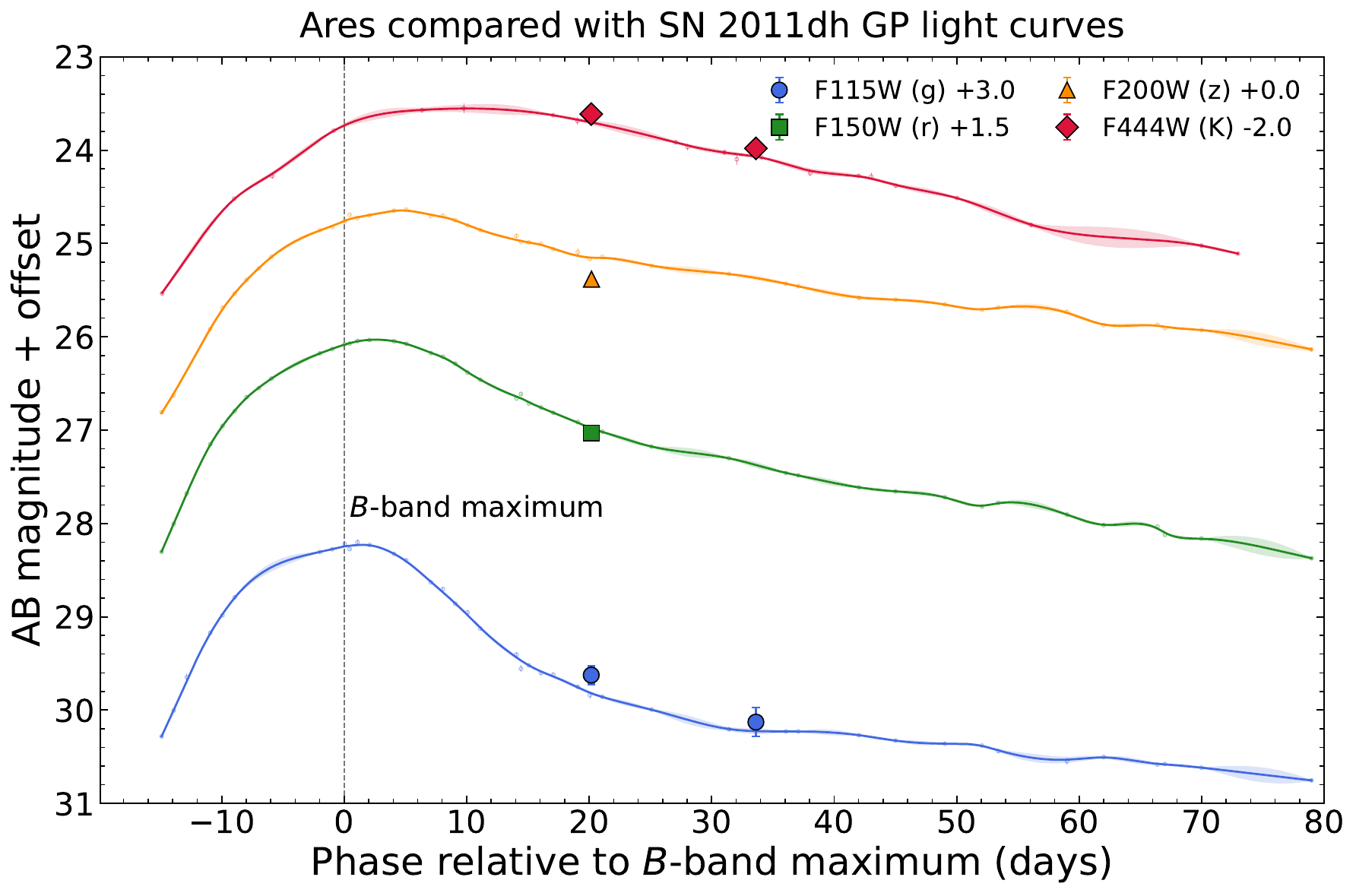}
    \includegraphics[width=0.49\linewidth]
        {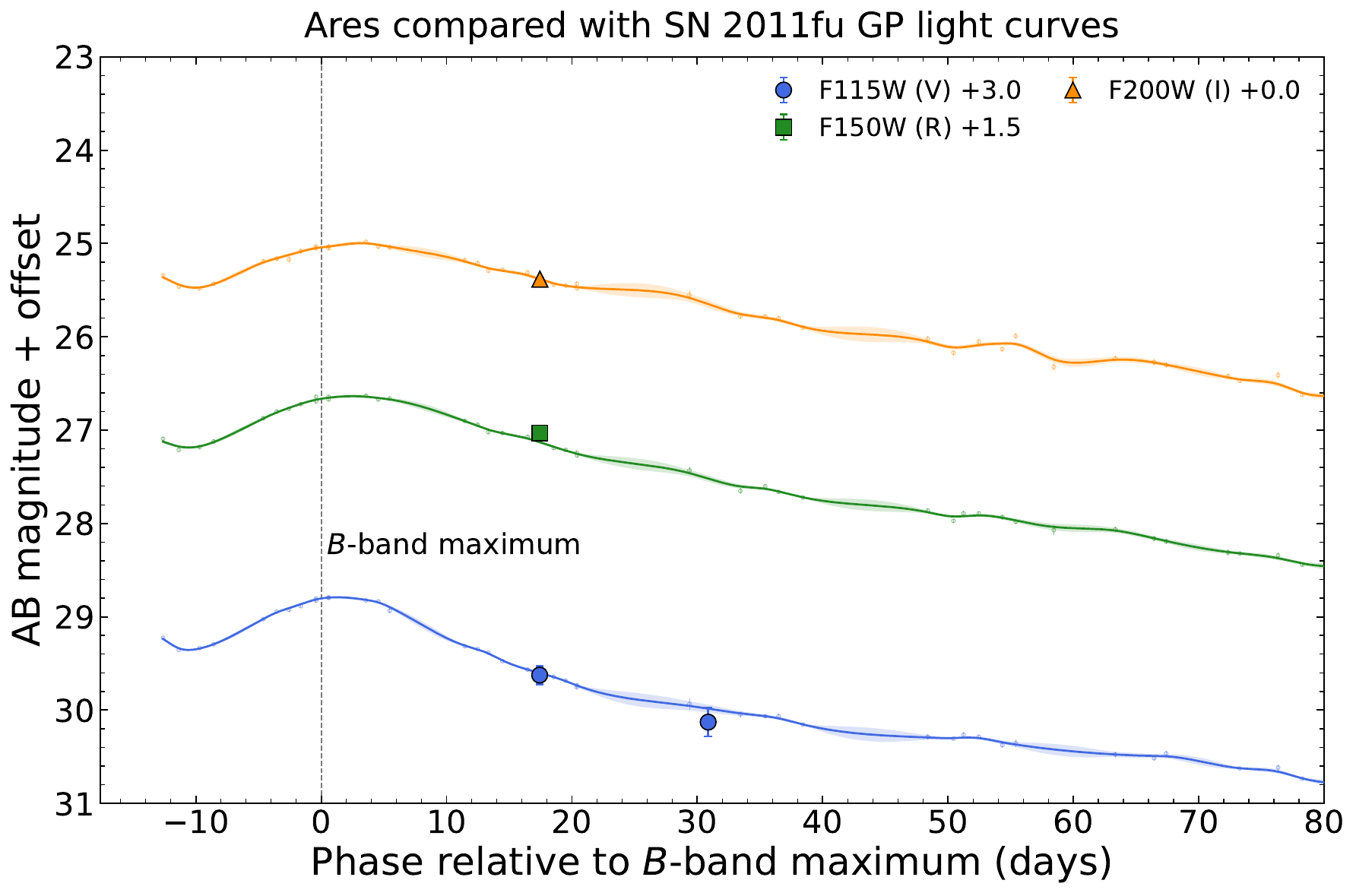}
    \caption{
    Supplementary light-curve comparisons using SN~2011dh (left)
    and SN~2011fu (right). Large filled symbols show the
    extinction-corrected SN~Ares photometry at the posterior-median
    phase of each fit. Curves and shaded bands show the shifted
    comparison-SN GP predictions and their $1\sigma$ uncertainties;
    small open symbols show the comparison-SN photometry.
    Filter mappings and visual magnitude offsets are indicated
    in the legends. The dashed vertical lines mark the radioactive
    $B$-band maximum of each comparison template.
    }
    \label{fig:additional_lc_fits}
\end{figure*}

We present supplementary figures for the light-curve analysis described in Section~5.2. Figure~\ref{fig:additional_lc_fits} shows the fits using SN~2011dh and SN~2011fu, complementing the SN~1993J and population-level comparisons shown in Figure~6. Both comparisons place the second \textit{JWST} epoch approximately one month after the radioactive $B$-band maximum, consistent with the main-text results.

\begin{figure*}[!h]
    \centering
    \includegraphics[width=0.49\linewidth]
        {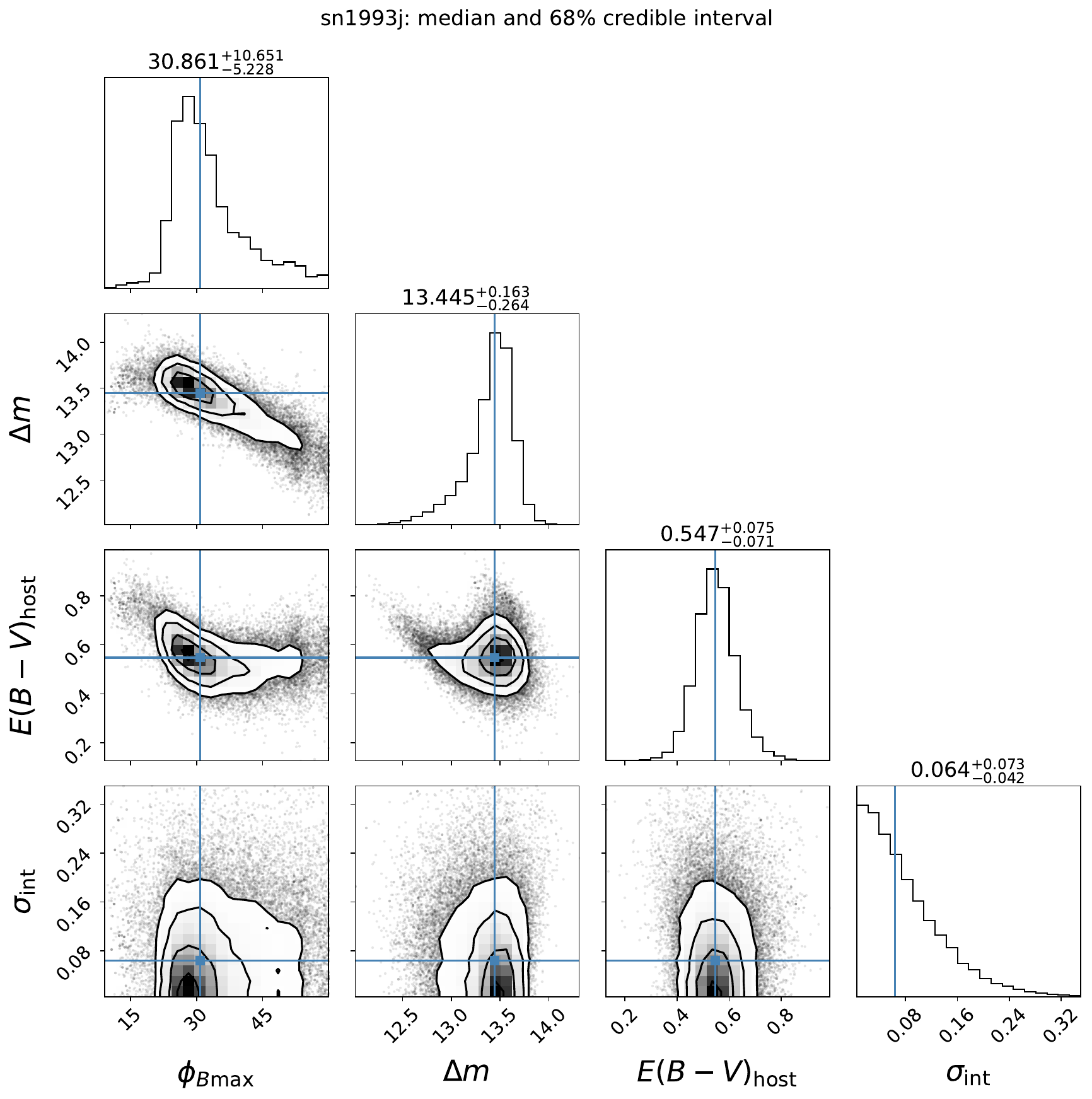}
    \includegraphics[width=0.49\linewidth]
        {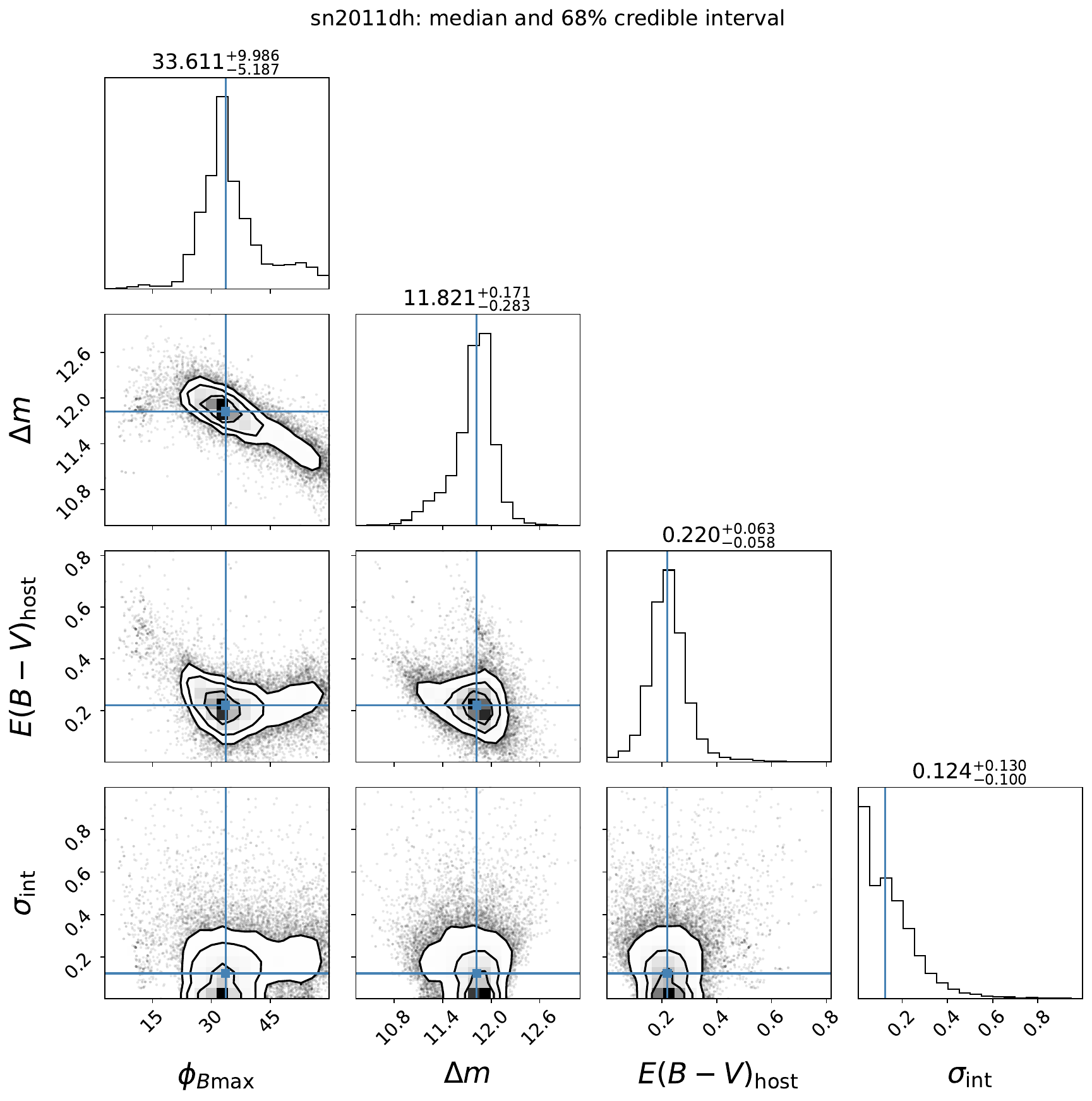}
    \caption{
    Posterior distributions for the SN~1993J (left) and
    SN~2011dh (right) light-curve comparisons.
    The parameters are the rest-frame phase relative to radioactive
    $B$-band maximum at the second \textit{JWST} epoch,
    $\phi_{B\max}$; the common magnitude offset, $\Delta m$;
    the SN~Ares host-galaxy color excess, $E(B-V)_{\rm host}$;
    and the intrinsic scatter, $\sigma_{\rm int}$.
    Blue lines indicate the marginal posterior medians.
    Values above the diagonal panels give the medians and
    uncertainties defined by the 16th and 84th percentiles.
    }
    \label{fig:corner_individual}
\end{figure*}

\begin{figure*}[!h]
    \centering
    \includegraphics[width=0.49\linewidth]
        {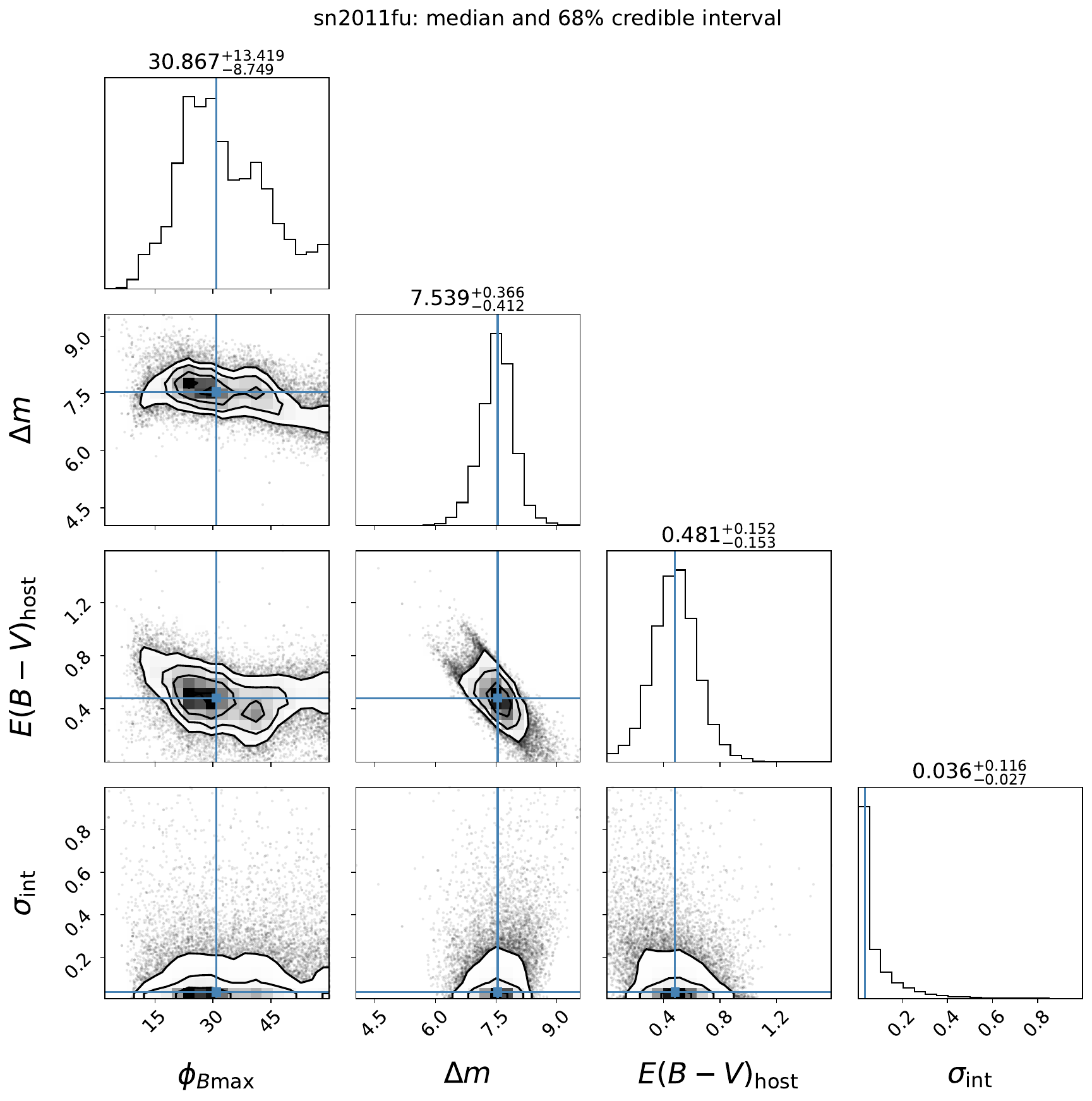}
    \includegraphics[width=0.49\linewidth]
        {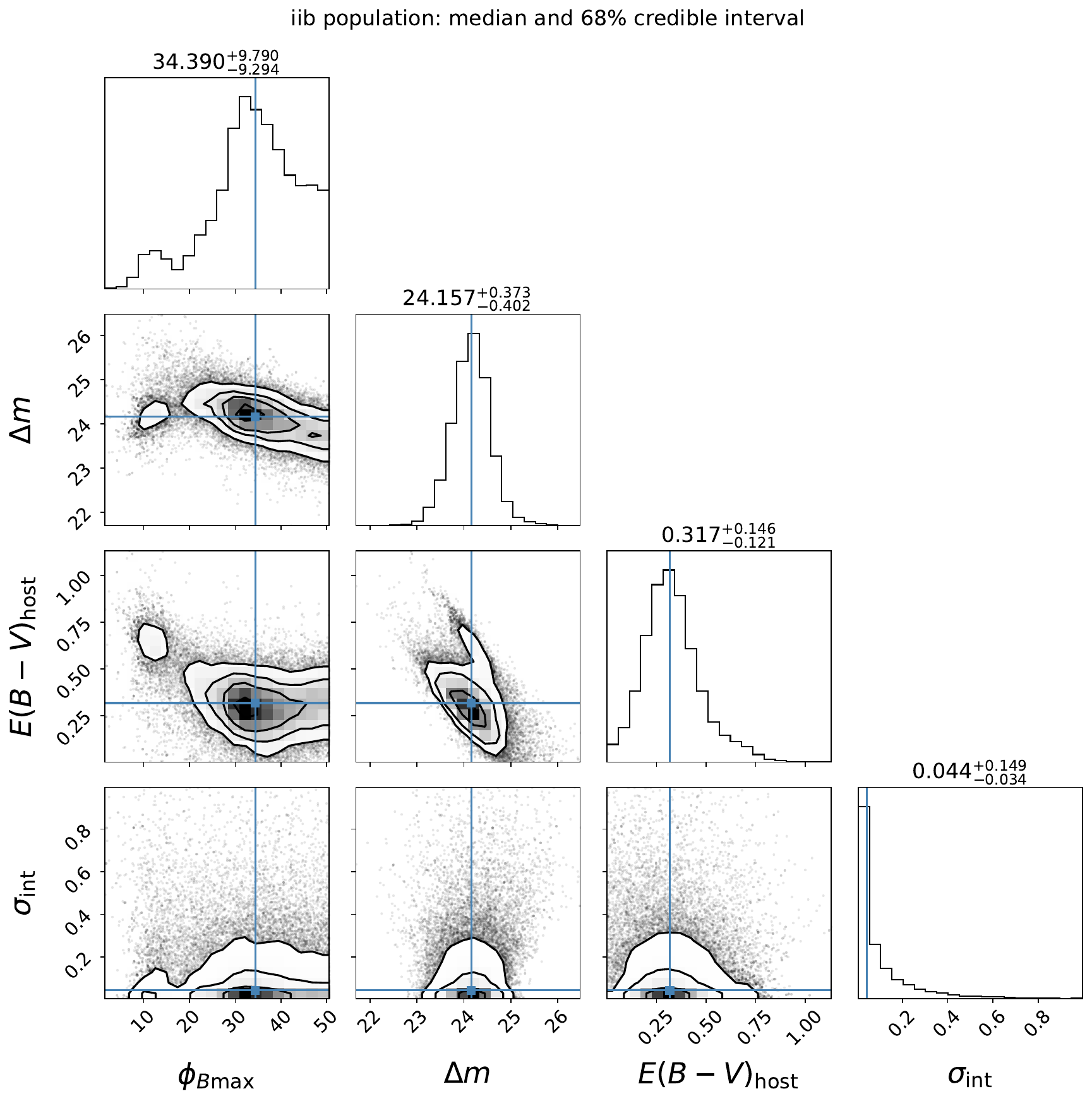}
    \caption{
    Posterior distributions for the SN~2011fu (left) and
    population-level SN~IIb (right) comparisons.
    Parameter definitions and plotting conventions are the same
    as in Figure~\ref{fig:corner_individual}.
    The population comparison includes a secondary low-phase
    solution associated with higher host-galaxy reddening.
    }
    \label{fig:corner_fu_population}
\end{figure*}

Figures~\ref{fig:corner_individual} and~\ref{fig:corner_fu_population} show the posterior distributions for all four comparisons. The diagonal panels display the marginalized distributions of phase, magnitude offset, host-galaxy reddening, and intrinsic scatter; the off-diagonal panels show their joint distributions. The phase distributions exhibit asymmetric tails and, for the population comparison, a secondary low-phase mode correlated with higher host reddening. These features illustrate the degeneracies permitted by the available photometry. The quoted uncertainties in the main text summarize the central 68\% credible intervals.

\clearpage

\bibliography{references,helios_refs}{}
\bibliographystyle{aasjournalv7}



\end{document}